\documentclass{article}
\usepackage{cancel}
\usepackage{hyperref}
\usepackage{xspace}
\usepackage{graphicx} 
\usepackage{float} 
\usepackage{todonotes}
\usepackage{amsmath}
\usepackage{caption}
\usepackage{subcaption}
\usepackage[percent]{overpic}
\usepackage{xcolor}
\usepackage[utf8]{inputenc}

\title{Gyrokinetic modelling of anisotropic energetic particle driven instabilities in tokamak plasmas}
\author{B. Rettino$^1$, T. Hayward-Schneider$^1$, A. Biancalani$^{2,1}$, A. Bottino$^1$,\\ Ph. Lauber$^1$, I. Chavdarovski$^3$, F. Vannini$^1$, F. Jenko$^1$ }
\date{%
    $^1$Max-Planck-Institut f\"ur Plasmaphysik, 85748 Garching, Germany\\%
    $^2$L\'eonard de Vinci P\^ole Universitaire, Research Center, 92916 Paris la D\'efense, France\\%
    $^3$Korea Institute of Fusion Energy, 34133 Daejeon, South Korea\\[2ex]%
}

\begin{document}

\maketitle

\begin{abstract}
    Energetic particles produced by neutral beams are observed to excite energetic-particle-driven geodesic acoustic modes (EGAMs) in tokamaks.  We study the effects of anisotropy of distribution function of the energetic particles on the excitation of such instabilities with ORB5, a gyrokinetic particle-in-cell code.  
    Numerical results are shown for linear electrostatic simulations with ORB5.  The growth rate is found to be sensitively dependent  on  the  phase-space  shape  of  the  distribution  function.  The behavior of the instability is qualitatively compared to the theoretical analysis of dispersion relations. Realistic neutral beam energetic particle anisotropic distributions are obtained from the heating solver RABBIT and are introduced into ORB5 as input distribution function. Results show a dependence of the growth rate on the injection angle. A qualitative comparison to experimental measurements is presented and few disagreements between them are found, being the growth rate in the simulations much lower than that from experiments. An explanation for the difference is advanced.

\end{abstract}

\section{Introduction}

The geodesic acoustic mode (GAM) is a toroidally symmetric (n=0) acoustic perturbation of density and electric potential in toroidal plasma configurations. GAMs were theoretically first described by Winsor et al \cite{winsor1968geodesic} through an ideal MHD approach, giving proper estimate to the mode frequency, but excluding the kinetic effects. This first derivation was useful to estimate the frequency of the mode but missing kinetic effects, such as the Landau damping, could not describe the physics of the interaction with the particles and a more accurate estimation of the damping rate.  Later, its dispersion relation was derived from a gyrokinetic approach \cite{smolyakov2016dispersion,zonca1996kinetic}. The GAM is found to be stable due to collisionless Landau damping with thermal ions \cite{sugama2006collisionless,qiu2008collisionless}. Particularly, the mode is found to be localized near the edge of the plasma column, since the damping rate is inversely proportional to $q$ \cite{qiu2008collisionless}, where $q$ is the tokamak safety factor. Energetic particles (EPs) can excite this mode, generating an energetic-particle driven GAM or EGAM \cite{qiu2010nonlocal,fu2008energetic,chavdarovski2021linear,novikau2019implementation,nazikian2008intense,zhiyong2011kinetic}. The mode can be excited by inverse Landau damping with the resonant EPs, which causes a redistribution of the latter from higher to lower energies \cite{biancalani2017saturation}.
Linear excitation of EGAMs with both circulating energetic particles \cite{qiu2010nonlocal,chavdarovski2021linear}, as well as trapped particles have been studied \cite{chavdarovski2021linear}, using the deeply trapped particle model  \cite{chavdarovski2009effects,chavdarovski2014analytic}. Additionally, theory of EGAMs excited by not fully slowed-down energetic ions was derived in Ref.\cite{cao2015fast} in reference to experimental observations
in Large Helical Device \cite{ido2015identification}. 
The mechanism of wave-particle energy exchange works with opposite contributions for EPs (giving energy to the mode, exciting it) and thermal ions (absorbing energy from it, damping it), therefore EGAMs are shown to be an effective means to transfer energy from high energy species to colder ones \cite{novikau2019implementation,doi:10.1063/1.5142802}. Experimental evidence  showed that EGAM's frequency is about $50\%$ of the usual GAM frequency\cite{nazikian2008intense}. Such observations have been supported by analytical computations of GAM dispersion relation and confirmed through numerical simulations \cite{zarzoso2014analytic,girardo2014relation,zarzoso2012fully,novikau2019implementation}. Nevertheless, it has been also shown that various unstable branches exist at different frequencies \cite{girardo2014relation}.

Plasma parameters can affect GAM/EGAM's growth rate. EP density is one of the main parameters affecting it. Simulations and analytical computations showed that an increase in EP concentration leads to an increase of growth rate of the mode, marking at a certain value of energetic particle fraction with respect to electron density $n_{EP}/n_e$ the threshold value for the transition from a damped to an excited mode \cite{qiu2010nonlocal,fu2008energetic,chavdarovski2021linear,di2018effect,zarzoso2012fully,zarzoso2014analytic,girardo2014relation}. On the other hand, the frequency is usually found to be decreasing because of the transition from an higher frequency GAM to a lower frequency EGAM for increasing EP fraction. As already mentioned, the safety factor profile $q$ is affecting the GAM growth rate determining the radial position at which the mode will be located \cite{qiu2008collisionless,biancalani2017cross}. Plasma elongation was theoretically predicted to affect the growth rate \cite{gao2010plasma}, while numerical simulations eventually verified the theory \cite{di2018effect,biancalani2017cross}.

The EP distribution function has a great influence on EGAM's growth rate. Firstly, an anisotropy in velocity space is necessary for the excitation of the mode \cite{betti1992stability}. Furthermore, negative gradients in kinetic energy in the distribution function ($\partial f/\partial E<0$) lead to damping of the EGAM \cite{todo2019introduction}. Whereas, positive gradients of the distribution functions in velocity space are needed to drive EGAMs, and as a consequence of which a certain portion of EPs will be redistributed to lower energies. \cite{biancalani2018nonlinear}.
In previous work, analytical anisotropic distribution functions are bump-on-tail \cite{novikau2019implementation,biancalani2018nonlinear,chavdarovski2021linear} and slowing down with pitch dependency \cite{qiu2010nonlocal,zhiyong2011kinetic,zarzoso2012fully,chavdarovski2021linear}, as well as single pitch Maxwellian \cite{chavdarovski2021linear}. Analytical theory and simulations adopting pitch dependent slowing down distribution function described it as function of $\Lambda=\mu B/E$, with $\mu$ being the magnetic moment of the particle, $E$ total kinetic energy and $B$ background magnetic field. 

The aim of this article is to study the effects of realistic distribution functions in order to obtain fully predictive simulations of experimental scenarios. 
For this purpose a new distribution function is formulated and implemented in the gyrokinetic code ORB5 \cite{lanti2020orb5} (Section \ref{secf0}). The effects of the shape of the distribution function in the velocity space, in terms of its parameters, are reported and discussed (in Section \ref{secorb5}). The numerical results are supported with theoretical analysis of the newly implemented distribution function, following the dispersion relation studies carried out in \cite{qiu2010nonlocal,chavdarovski2021linear,girardo2014relation} (Section \ref{secanalysis}). Finally, experimental relevant distributions functions from Fokker-Planck solver code RABBIT  \cite{Weiland_2018} are presented and later used to obtain realistic simulations of NLED-AUG case \cite{vannini2020gyrokinetic} with experimental density and temperature profiles. Results are qualitatively compared with NLED-AUG case experimental data \cite{lauber2014off} and then discussed (in Section \ref{secrabbit}).


\section{Theoretical model}
\label{secf0}

\subsection{Vlasov equation in ORB5}
The code ORB5 solves the gyrokinetic (GK) Vlasov equation, coupled with relevant GK field equations, typically a Polarization equation (Poisson) and, in the electromagnetic model,  a parallel Amp\`eres law \cite{lanti2020orb5}.
The GK Vlasov equation for the particle species $p$, in the absence of collisions and sources, reads:
\begin{align}
    \frac{d f_p}{d t}&=0~,
\end{align}
$d/dt$ is the convective derivative. The full derivation of the GK model of ORB5 can be found in \cite{tronko2016second,tronko2017hierarchy}. The distribution function is then decomposed into an analytically known background $f_{0}$, solution of the unperturbed Vlasov equation, and a perturbed distribution functions $\delta f$. The Vlasov equation becomes now an evolution
equation for $\delta f$
\begin{equation}
\frac{d \delta f}{d t}=-\frac{d f_0}{d t}~,
\end{equation}
where $f_0$ is typically written as a function of the kinetic energy, the adiabatic invariant per unit mass $\mu$ and the gyrocenter position $\textbf{R}$.

\subsection{Analytical slowing down with pitch dependency}
As explained in the introduction distribution functions have a strong impact on the GAM behaviour. It is of interest to study how different shapes in phase space influence the stability of the modes.
To this purpose, we have implemented a new analytical distribution function for a pitch-angle dependent slowing down particles to compare analytical distribution function results with experimental ones, namely using distribution functions from RABBIT \cite{Weiland_2018}.
The distribution function is a function of energy and parallel velocity, both normalized with respect to the sound speed $v_{s}=\sqrt{T_e/m_i}$ (with $k_B$, Boltzmann constant, understood), where $T_e$ is the electron temperature and $m_i$ the ion mass:

\begin{equation}
\hspace*{-1.5cm}
    f(v,\xi,\psi)=n_{val}(\psi)\frac{2\sqrt{\frac{2}{\pi}}}{\sigma_{\xi}[erf(\frac{\xi_0+1}{\sqrt{2}\sigma_{\xi}})-erf(\frac{\xi_0-1}{\sqrt{2}\sigma_{\xi}})]}\exp\left(-\frac{(\xi-\xi_0)^2}{2\sigma_{\xi}^2}\right)\frac{3\; \Theta(v_{\alpha}-v)}{4\pi (v_c^3(\psi)+v^3)}ln\left(1+(\frac{v_{\alpha}}{v_c(\psi)})^3\right)
    \label{Eq1.a}
\end{equation}

The analytical distribution function was obtained as a product of a slowing down in energy, characterized by the absolute value of velocity $v=\sqrt{2\varepsilon}$ (3$^{rd}$ and 4$^{th}$ RHS terms of Eq.\ref{Eq1.a}), and a gaussian in $\xi=v_{\parallel}/|v|$ ($\xi$ can range only from -1 to 1) centered in $\xi_0$ (1$^{st}$ and 2$^{nd}$ RHS terms in Eq.\ref{Eq1.a}), characterized by a width $\sigma_{\xi}$. The distribution function depends radially on the magnetic flux coordinate $\psi$, normalized with respect to its value at the separatrix. Here, $\Theta(v_{\alpha}-v)$ is the Heaviside function defined as 1 for values of $v<v_{\alpha}$ and 0 elsewhere, where $v_{\alpha}$ is the injection velocity, and $v_c(\psi)$ is the critical velocity of plasma, calculated from bulk plasma parameters (especially $T_e$) hence depending on the radial coordinate $\psi$; $n_{val}(\psi)$ is the local value of the normalized density. This distribution function was constructed on purpose to both create a versatile tool which could be analyzed in function of its two main parameters ($\xi_0$ and $\sigma_{\xi}$) and to produce an analytical distribution function which would best resemble the experimental distribution functions obtained from the Fokker-Planck solver code RABBIT \cite{Weiland_2018} (see Fig.\ref{fig9.a}). 
Note that this distribution function is not a solution of the equilibrium Vlasov equation, due to the explicit dependence on the parallel velocity. The secular evolution of the distribution function on the unperturbed trajectories has been neglected in the simulations to be consistent with the experiments, in which particle and heat sources are present. This distribution function represents a further step in the direction of the experimental-like distribution functions with respect to the isotropic slowing down presented in \cite{vannini2021studies}.

\begin{figure}[H]
\hspace{1cm}
    \begin{overpic}[width=10cm,tics=10]{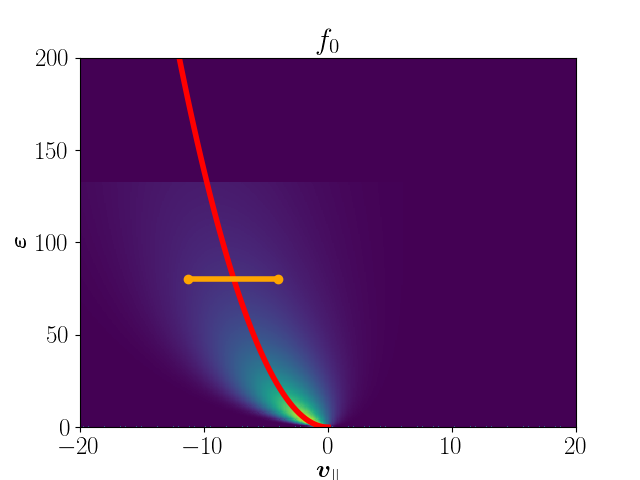}
        \put (17,60) {\color{red}\huge$\displaystyle\xi_0$}
        \put (45,40) {\color{orange}\huge$\displaystyle\sigma_{\xi}$}
    \end{overpic}
    \caption[width=5cm,height=5.0cm]{Analytical slowing down with pitch angle dependency ($\xi_0=-0.6,\sigma_{\xi}=0.4$), on the right $\xi_0$ and $\sigma_{\xi}$ are qualitatively represented in the $f_0$ plot}
    \label{fig1.a}
\end{figure}


ORB5 distribution functions are represented in the phase space used in Fig.\ref{fig1.a}. The two coordinates used are parallel velocity $v_{\parallel}$, normalized with respect to the sound velocity $v_{s}=\sqrt{T_e/m_i}$, and a normalized energy $\varepsilon=v_{\parallel}^2/2+\mu B$. All the points lying under the $\mu=0\implies \varepsilon=v_{\parallel}^2/2$ parabola are non physical, hence the code will not
consider them. It is interesting to note the intrinsic difference between $v_{\parallel}$ and $\xi$. The first parameter is normalized with respect to a constant value $v_{s}$, while the latter is normalized with respect to the modulus of velocity which varies with energy. This causes the characteristic curvature in the analytical slowing down with pitch dependency and implies a reduction in the "width" of $f_0$, measured in $v_{\parallel}$ units, as we move to lower values of energy (or equivalently $|v|$).

Once defined the new distribution function, we show the results obtained from ORB5 simulations. The effects of the new distribution function are analyzed in function of the two main parameters which characterize the new slowing down distribution of energetic particles: $\xi_0$ and $\sigma_{\xi}$.

\section{Numerical results from NLED-AUG case}
\label{secorb5}
\begin{figure}[H]
    \includegraphics[width=5cm,height=6cm]{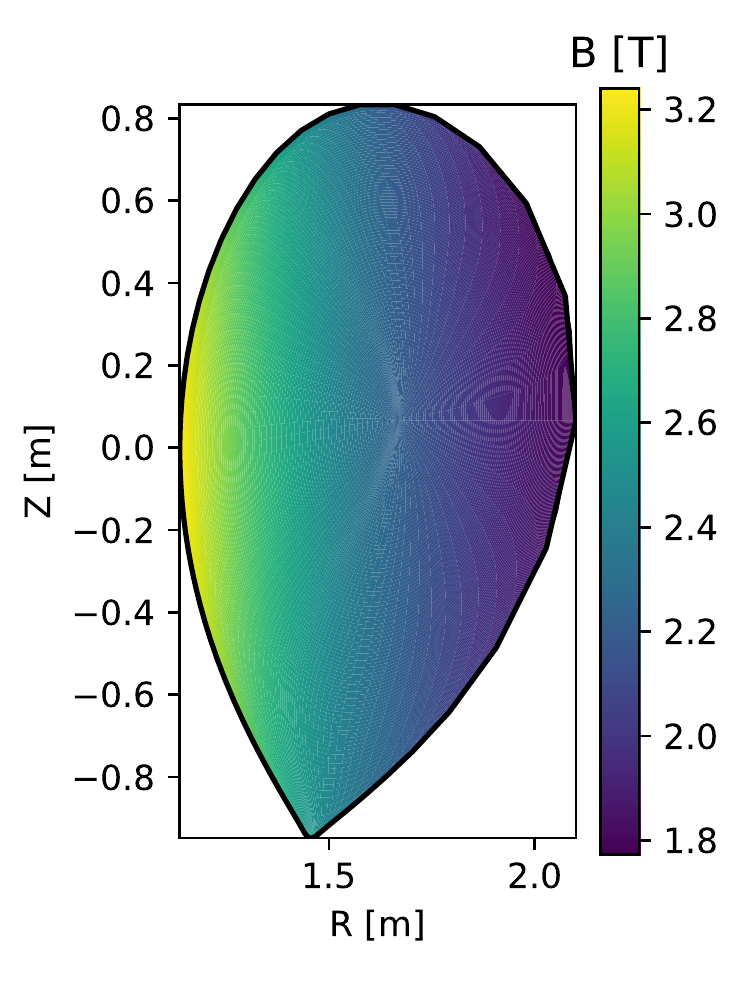}
    \includegraphics[width=6cm,height=5.5cm,tics=10]{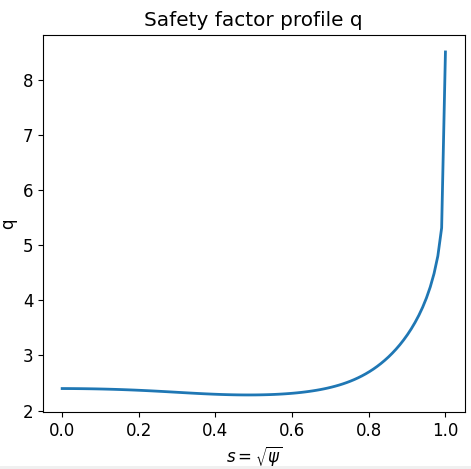}
    \caption[width=5cm,height=5.0cm]{NLED-AUG case magnetic equilibrium (left), safety factor profile (right)}
    \label{fig3.b}
\end{figure}

\begin{figure}[H]
\hspace{1cm}
    \includegraphics[width=10cm]{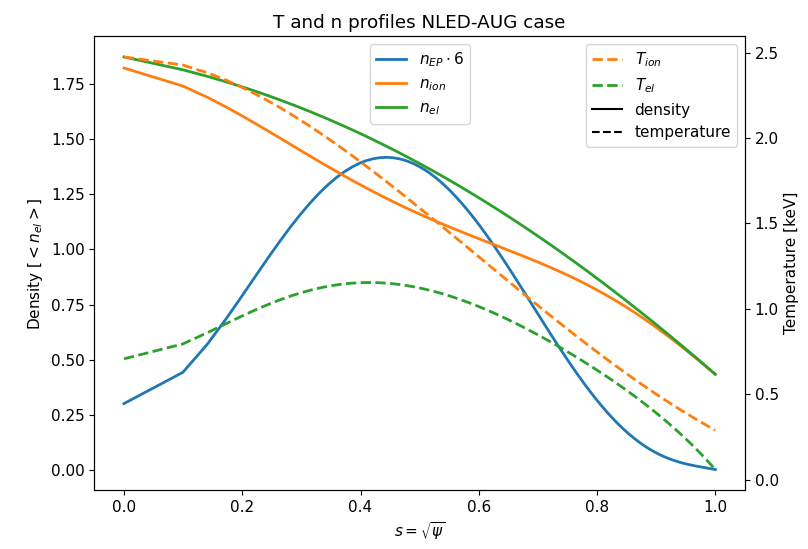}
    \caption[width=5cm,height=5.0cm]{NLED-AUG case temperature and density profiles \cite{vannini2020gyrokinetic}}
    \label{fig3.bcc}
\end{figure}

Next, we will describe the results of ORB5 simulations of the NLED-AUG case \cite{lauber2014off}, performed varying the parameters $\xi_0,\;\sigma_{\xi}\text{ and }n_{EP}$. Excitation threshold values will be found numerically for these parameters. NLED-AUG case is based on the ASDEX-Upgrade (AUG) shot 31213 at 0.84s. EPs were injected in the plasma at 93keV through an off-axis neutral beam injection (NBI) at an angle of 7.13$^{\circ}$ with respect to the magnetic axis. This shot is characterized by the magnetic equilibrium shown in Fig.\ref{fig3.b} (a), and an off-axis EP density profile as in Fig.\ref{fig3.bcc}. Some of the main parameters characterizing ASDEX-Upgrade shot 31213 are reported in Table \ref{tab1.a}. More details about NLED-AUG case can be found in reference \cite{Laubweb}.

\begin{table}[H]
\begin{tabular}{|c c c c c|} 
 \hline
 $a_0$ [m] & $R_0$ [m] & $B_0$ [T] & $\Omega_{ci}$ [rad/s] & $\omega_A$ [rad/s] \\ 
 \hline
 0.482 & 1.666 & 2.202 & 1.054e8 & 4.98e6\\
 \hline \hline
 $T_e|_{s=0}$ [keV] & $T_i|_{s=0}$ [keV] & $n_e|_{s=0}$ [$m^{-3}$] & $n_i|_{s=0}$ [$m^{-3}$] & $n_{EP}|_{s=0}$ [$m^{-3}$]\\
\hline
0.709 & 2.48 & 1.672e19 & 1.6018e19 & 6.98e17\\
\hline

\end{tabular} 
\caption{Parameters of NLED-AUG case \cite{vannini2020gyrokinetic}}
\label{tab1.a}
\end{table}

The threshold values which will be presented in the following sections are not directly comparable to RABBIT distribution function cases. In fact, RABBIT distribution functions are not directly related to some analytical case obtained from Eq. \ref{Eq1.a} for some value of its parameters, especially $\sigma_{\xi}$. RABBIT distributions present an isotropization effect going to lower energies, something missing in the analytical slowing down, how it is shown in Fig.(\ref{fig1.a}). Additionally, the analytical slowing down doesn't represent the other two injection velocities at $E/2$ and $E/3$ present in the RABBIT $f_0$. Nevertheless, looking at the RABBIT distribution functions, and keeping in mind the characteristic injection angles of NLED-AUG case, we could tell that the corresponding $\xi_0$ for the RABBIT $f_0$ $\sim-0.9$. Furthermore, the trends of growth rate with respect of the $f_0$ parameters that will be found in the next sections will be observed also in the linear simulations for the RABBIT distribution functions.

\subsection{Scan in $\xi_0$ and $\sigma_{\xi}$}

\begin{figure}[H]
    \centering
    \includegraphics[height=7cm]{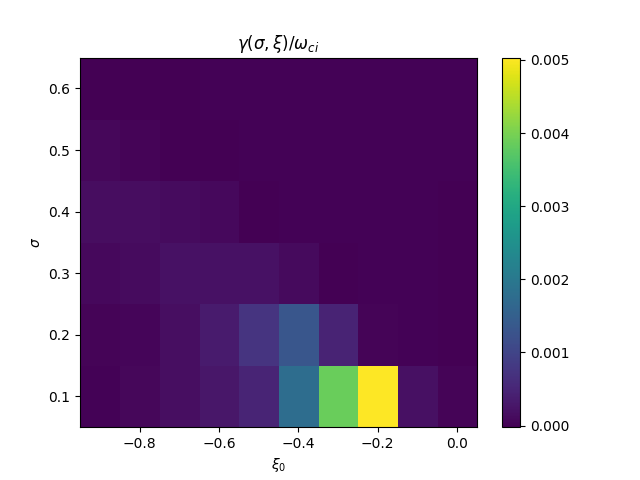}
    \caption{Growth rate as function of $\xi_0$ and $\sigma_{\xi}$ (NLED-AUG case), on the right top-view of the left plot}
    \label{fig5.a}
\end{figure}

We present in this section a scan in both $\xi_0$ and $\sigma_{\xi}$ of the EGAM growth rate. The scan has been obtained by running NLED-AUG case, described above, with ORB5. The code has been run linearly, electrostatically, considering adiabatic electrons \cite{lanti2020orb5,BottinoJPP2015}. The distribution function, used as input, was obtained implementing equation (\ref{Eq1.a}) picking $\xi_0$ and $\sigma_{\xi}$ from a mesh of these two parameters, with $\xi_0$ ranging between $0.0$ and $-0.9$ and $\sigma_{\xi}$ ranging between $0.1$ and $0.6$. The energetic particle concentration was constant for all the cases, namely $n_{EP}/n_e\simeq0.09$. The profiles are the same described above, namely those used in the off-axis case in reference \cite{vannini2020gyrokinetic}. Figure \ref{fig5.a} shows the qualitative behavior of the growth rate $\gamma$ as $\xi_0$ and $\sigma_{\xi}$ vary. Clearly, both parameters can affect $\gamma$. The global trend of the growth rate is decaying for distributions with larger $\sigma_{\xi}$, while it grows for values of $\xi_0$ included between 0.2 and 0.9. We notice that the maximum growth rate for every $\sigma_{\xi}$  shifts toward higher $\xi_0$ as $\sigma_{\xi}$ increases (Fig.\ref{fig5.a}). We actually notice that for high values of $\xi_0$, as $\sigma_{\xi}$ increases, we get a slight increase in growth rate before the isotropization effects prevail. This is not observed for others values of $\xi_0$, for which the decreasing trend of the growth rate with $\sigma_{\xi}$ is monotonic. As it will be explained later in Sec \ref{secmpr}, this may be due to the fact that for such high pitch angles, as $\sigma_{\xi}$ increases, the positive gradient of $f_0$ in parallel velocity move in the area where the most unstable modes get most of their drive. This could cause the excitation of the modes with high $\xi_0$ and $\sigma_{\xi}\sim0.4$ too. Anyhow, as $\sigma_{\xi}$ tends to $\infty$, the isotropization effects prevail and all the modes get damped. In the following subsections we will focus on the effects of the three parameters $\sigma_{\xi}$, $\xi_0$ and $n_{EP}$, trying to give threshold values for each of them fixing the other two parameters to some reference values.

\subsection{Effects of $\sigma_{\xi}$ and threshold value}

\begin{figure}[H]
    \includegraphics[width=6cm,height=5cm]{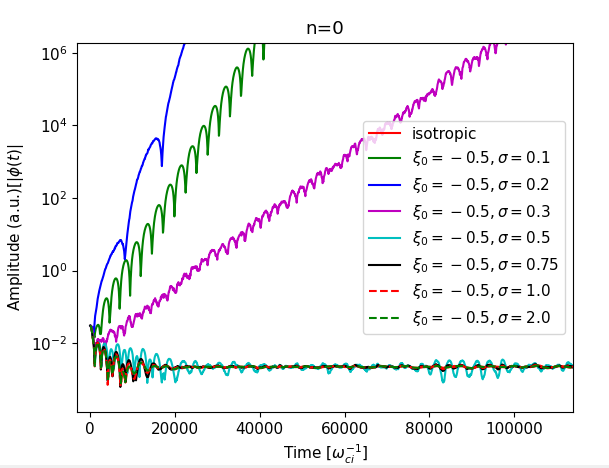}
    \includegraphics[width=6cm,height=4cm,tics=10]{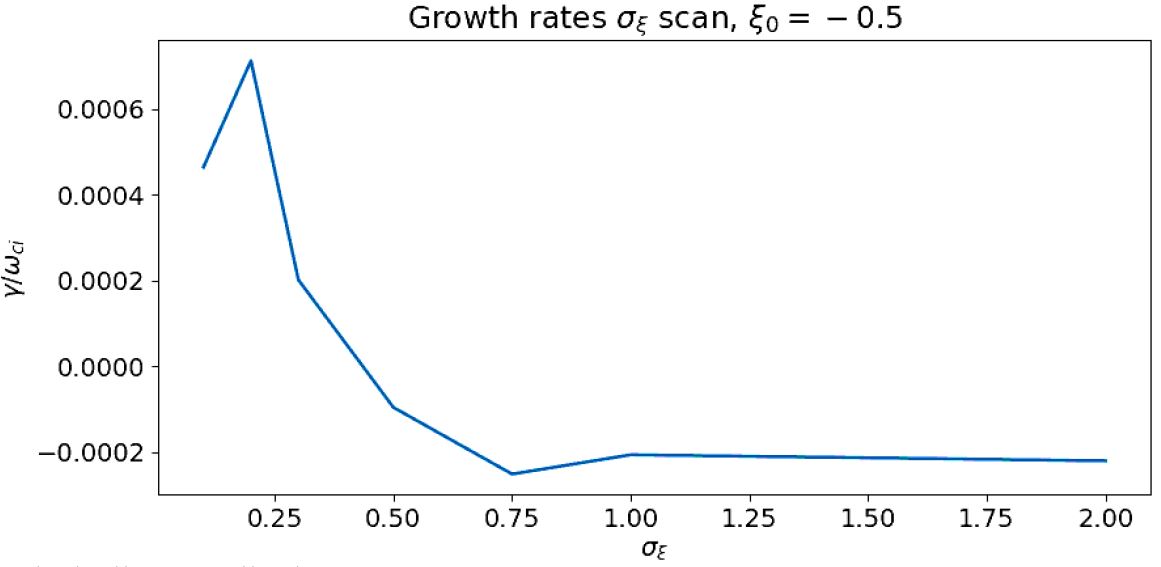}
    \caption[width=5cm,height=5.0cm]{$\sigma$ scan for $\xi_0=-0.5$ in NLED-AUG case. Plot on the left shows the amplitudes of the scalar potential field in time. Plot on the right shows the growth/damping rate dependence on $\sigma$}
    \label{fig2.a}
\end{figure}

It has been hypothesized that isotropic distribution functions in $v_{\parallel}$, provided they have everywhere negative gradients in energy ($\partial f/\partial \varepsilon<0$), cannot excite $n=0$ modes, namely EGAMs \cite{betti1992stability}. 
If $\sigma_{\xi}$ approaches $\infty$, pitch-dependent slowing down (SDs) become equivalent to isotropic SDs. It has been found that from values of $\sigma_{\xi}\sim0.5$ any pitch dependent SD wouldn't excite any EGAM, the mode (a GAM in this case) would rather be damped. Such results are shown in Fig.\ref{fig2.a}, where NLED-AUG case \cite{vannini2020gyrokinetic} simulations were run with the analytical slowing down with pitch dependency for $\xi_0=-0.5$ in a range of different $\sigma_{\xi}$. From Figure \ref{fig2.a}(a), representing the modes' radial infinity norm of scalar potential amplitude signals, we see that the isotropic slowing down and those, whose $\sigma\longrightarrow\infty$, don't trigger any EGAM. On the other hand, small values of $\sigma_{\xi}$ are able to drive EGAMs with higher growth rates since the gradients of $f_0$ are steeper. In conclusion, we can  state that the growth rate quickly decreases and eventually becomes negative as $\sigma_{\xi}$ values approach $\infty$. This trend is particularly clear in  Fig.\ref{fig2.a}(b), where the growth rates has been reported as function of the standard deviation $\sigma_{\xi}$.  

We found the excitation threshold value at $\sigma_{\xi}\simeq0.4$, considering $\xi_0=-0.5 \text{ and } n_{EP}/n_e=0.09$. As we can observe from Fig.\ref{fig5.a}, this value is indicative, since the highest growth rates are found for mid-range $\xi_0$. So the threshold value found for  $\xi_0=-0.5$ can be considered as a limit beyond which modes are stable for any $\xi_0<0.75$. As observed in the previous section high values of $\xi_0$ show a non-monotonic trend with $\sigma_{\xi}$ and therefore have slightly higher threshold values for $\sigma_{\xi}$. Anyhow as $\sigma_{\xi}\longrightarrow\infty$, the modes are damped because of the isotropization of the distribution function.

This result will find a theoretical confirmation in Section \ref{secanalysis}. 

\subsection{Effects of $\xi_0$ and threshold values}

Next, a scan has tested pitch-dependent slowing down with constant $\sigma=0.25$ over $\xi_0$ ranging from 0 to -0.9. The results are plotted below:

\begin{figure}[H]
    \centering
    \includegraphics[height=5.5cm]{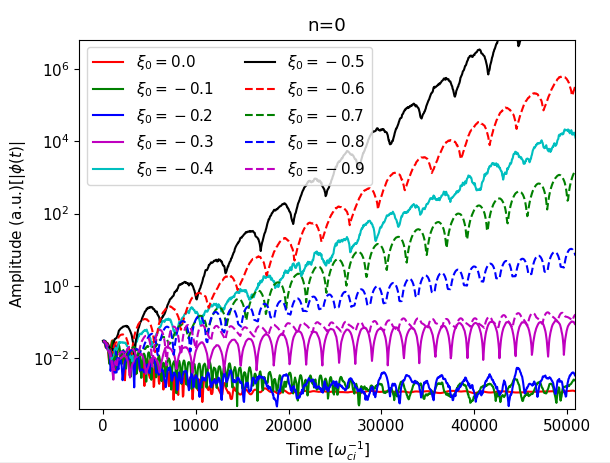}
    \caption{Mode amplitudes of a scan in $\xi_0$, ranging from 0.0 to -0.9, at $\sigma=0.25$ for NLED-AUG case}
    \label{fig.3.a}
\end{figure}

\begin{figure}[H]
    \centering
        \begin{overpic}[width=10cm,tics=10]{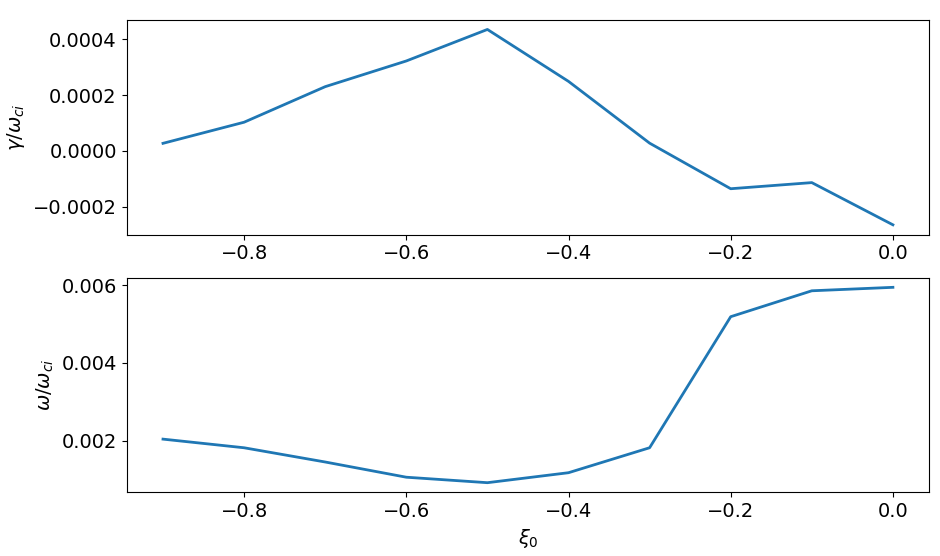}
        \put (-6,44) {\color{black}\Large$\displaystyle a)$}
        \put (-6,18) {\color{black}\Large$\displaystyle b)$}
    \end{overpic}
    \caption{Growth rate $\gamma$ (a) and frequency $\omega$ (b) of the scan in $\xi_0$ shown in Fig.\ref{fig.3.a}}
    \label{fig4.a}
\end{figure}

Both Figures \ref{fig.3.a} and \ref{fig4.a} show an evident increase of growth rate for mid-range values of $\xi_0$, while the mode is constant or damped for small or high values of $\xi_0$. References \cite{qiu2010nonlocal,zhiyong2011kinetic} predicted that  $\xi_0$ must be greater than a certain value to trigger an EGAM. Reference \cite{zarzoso2014analytic} instead predicted that both extremes values of $\xi_0$'s range have to be avoided in order to trigger an EGAM. Both references obtained such conditions in a different phase space, where the normalized parallel velocity was replaced by the perpendicular energy fraction $\Lambda$. Nevertheless, the results shown in Figure \ref{fig.3.a} and \ref{fig4.a} confirm the analytical predictions made so far. 

Theoretical analysis in Section \ref{secanalysis} will offer theoretical explanation to such behavior, completing the analytical description of the pitch effects obtained so far. Figure \ref{fig2.b}, plotting the integrand of Eq.\ref{eq8.a}, predicts the behavior of these numerical simulations in function of both parameters $\xi_0 \text{ and } \sigma_{\xi}$.

We also notice that, as the growth rate increases for mid-range values of $\xi_0$, there's a decay of frequency (Fig.\ref{fig4.a},b). The frequencies shown in the plot are Lorentzian fits of the frequency spectrum. In truth there are two modes (the GAM and the EGAM,  which differ by a factor 2, as said before) co-existing at the same time. For excited modes the EGAM contribution is greater, decreasing the Lorentzian fit value and vice versa.  As already discussed, this is also in accordance with theoretical, numerical and experimental evidence \cite{zarzoso2014analytic,girardo2014relation,nazikian2008intense}. 

Despite the excitation threshold is crossed at different $\xi_0$ values for different $\sigma_{\xi}$ and $n_{EP}/n_e$, from the results, plotted in Fig.\ref{fig5.a}, \ref{fig.3.a} and \ref{fig4.a}, it's clear that, whatever the other parameters are, the EGAM is excited for values of $\xi_0$ included in a certain interval $(\xi_{0,1},\xi_{0,2})$, where usually $-1\le\xi_{0,1}\le-0.5$ and $-0.5\le\xi_{0,2}\le0$. In our reference case, we fixed $\sigma_{\xi}=0.25$ and, as before, $n_{EP}/n_e=0.09$. For these parameters' values, the $\xi_0$ threshold interval is $(\xi_{0,1},\xi_{0,2})\simeq(-0.9,-0.3)$ (Fig.\ref{fig.3.a}).

\subsection{Scans in energetic particle concentration and threshold values}
\label{subsecfpartorb5}

\begin{figure}[H]
    \includegraphics[width=6cm,height=5cm]{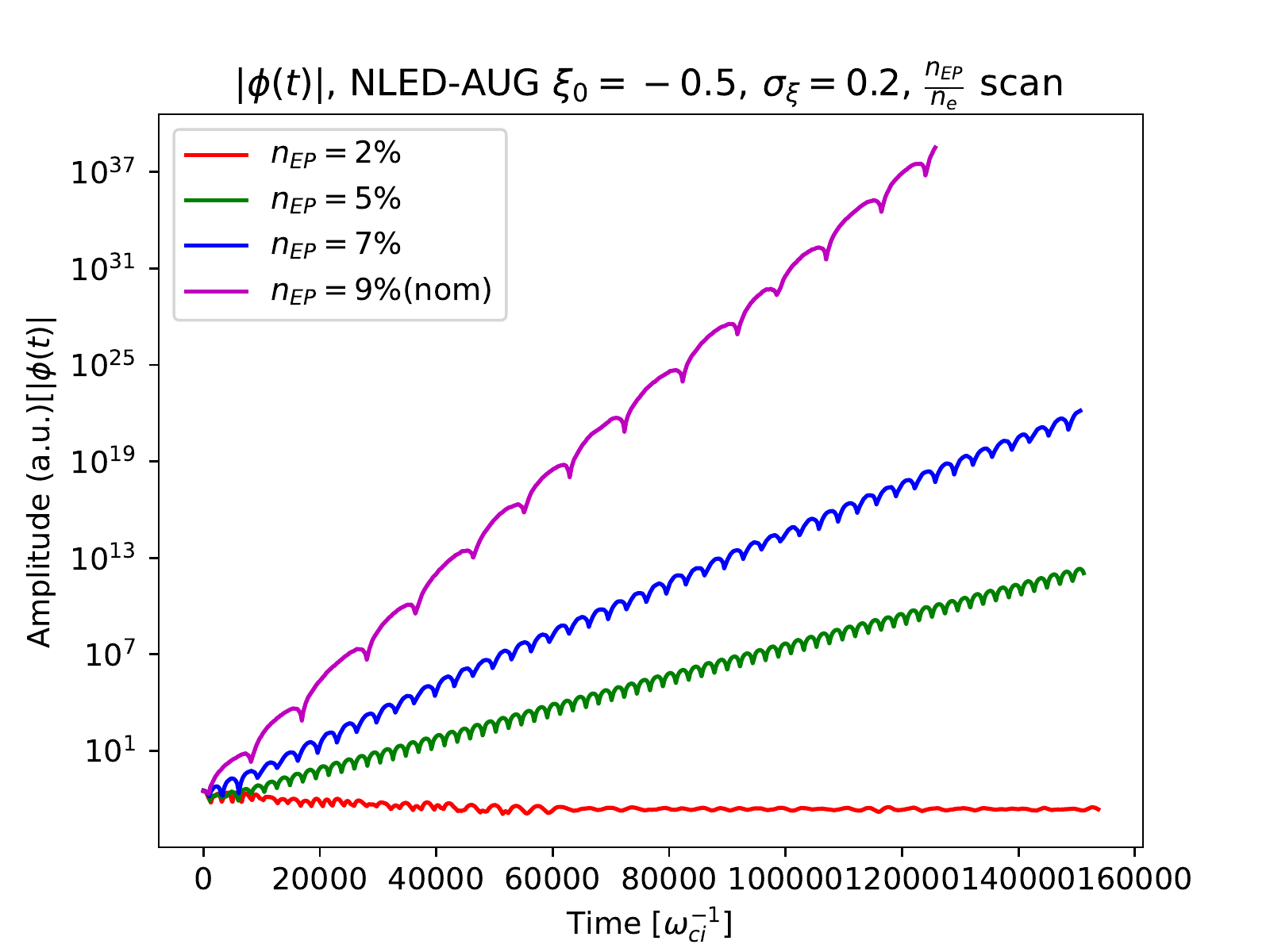}
    \includegraphics[width=6cm,height=5cm,tics=10]{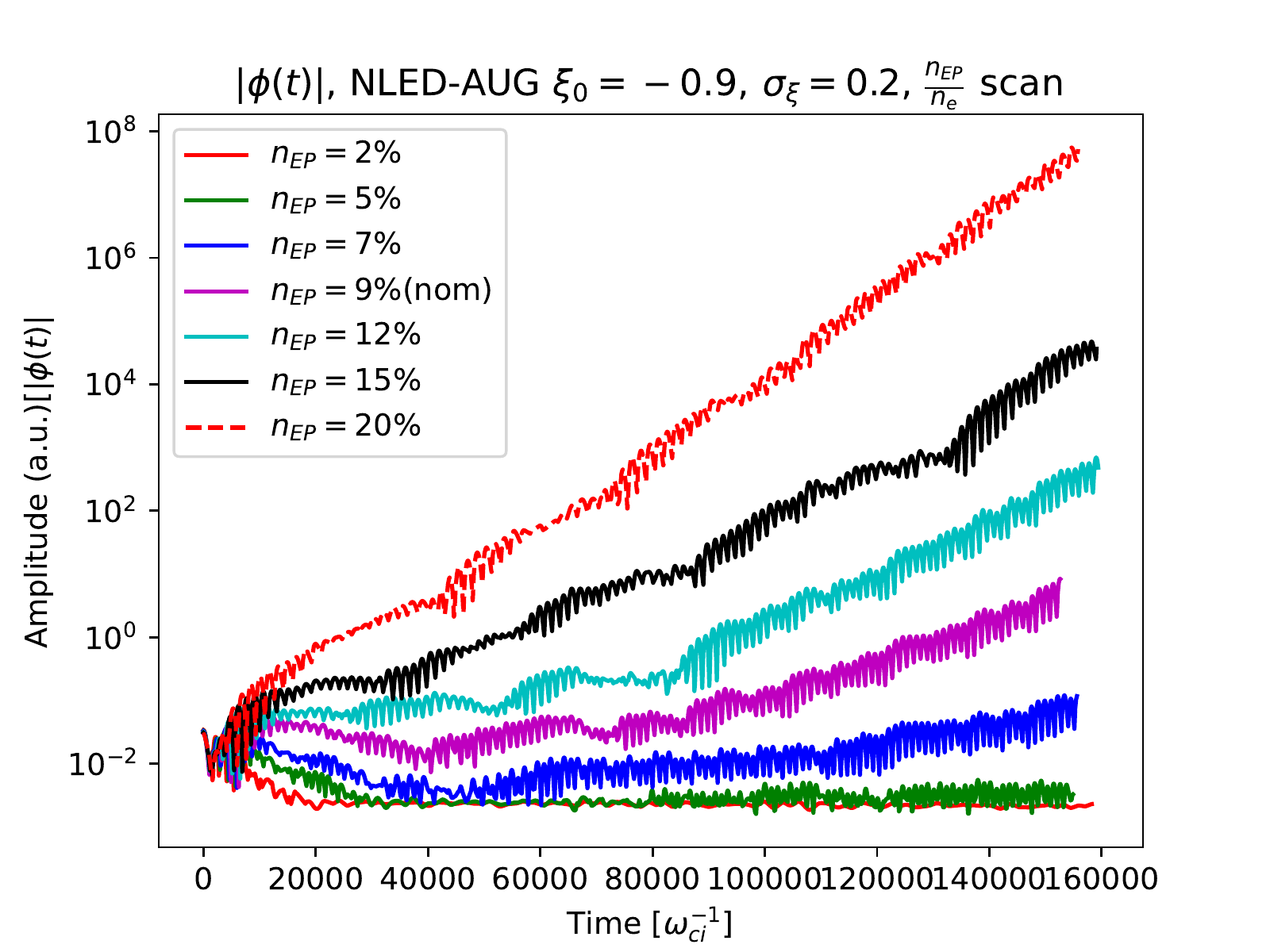}
    \caption[width=5cm,height=5.0cm]{$\frac{n_{EP}}{n_e}$ scan for $\xi_0=-0.5$ (left) and $\xi_0=-0.9$ (right) in NLED-AUG case}
    \label{fig8.b}
\end{figure}

\begin{figure}[H]
    \begin{overpic}[width=10cm,tics=10]{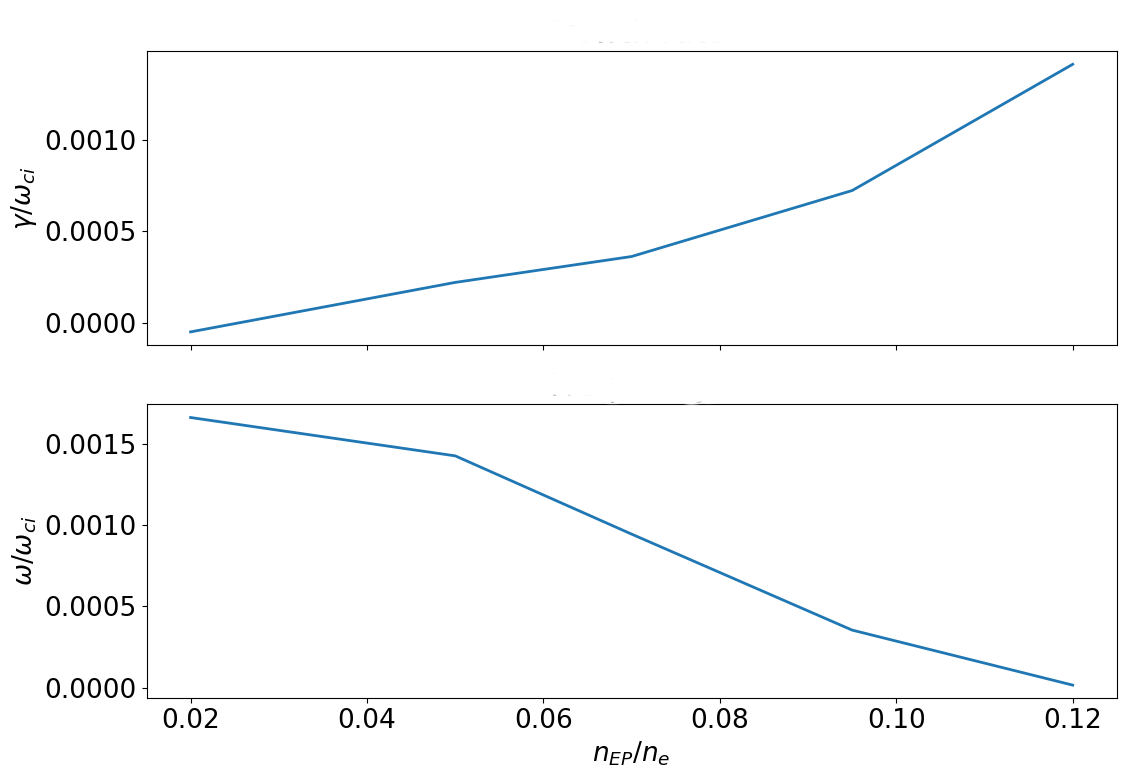}
        \put (-6,50) {\color{black}\Large$\displaystyle a)$}
        \put (-6,18) {\color{black}\Large$\displaystyle b)$}
    \end{overpic}
    \begin{overpic}[width=10cm,tics=10]{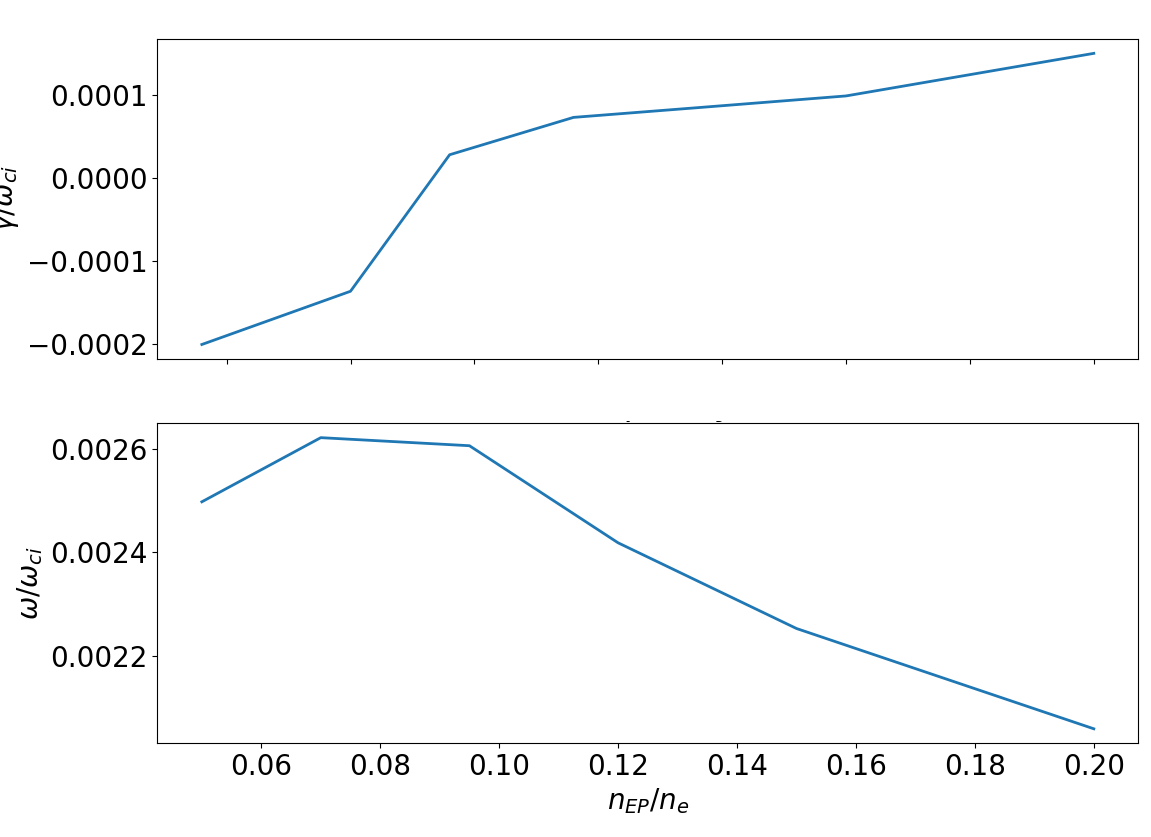}
        \put (-6,52) {\color{black}\Large$\displaystyle c)$}
        \put (-6,20) {\color{black}\Large$\displaystyle d)$}
    \end{overpic}
    \caption[width=5cm,height=5.0cm]{Growth rate and frequency dependence on $\frac{n_{EP}}{n_e}$ for $\xi_0=-0.5$ (a,b) and $\xi_0=-0.9$ (c,d) in NLED-AUG case}
    \label{fig9.b}
\end{figure}

In this section we scanned with respect to fast particle concentration two of the previous simulations, namely the one with $\xi_0=-0.5,\;\sigma_{\xi}=0.2$ and another one with $\xi_0=-0.9,\;\sigma_{\xi}=0.2$. Comparing the two scans, we can get an idea about how the $\xi_0$ parameter influences the threshold values of EP concentration. The scans have been run using again the NLED-AUG case configuration described at the beginning of this section. Figures \ref{fig8.b} and \ref{fig9.b} show the change in growth rate as we increase the fraction of EPs. For both $\xi_0=-0.5$ (Fig.\ref{fig8.b}, \ref{fig9.b} (left)) and  $\xi_0=-0.9$ ,(Fig.\ref{fig8.b},\ref{fig9.b} (right)) the growth rate $\gamma$ increases steadily with $n_{EP}/n_e$. But, as we can see from both figures below, the $\gamma=0$ threshold value is met at different values of $n_{EP}/n_e$ for the two cases. As expected, the case where $\xi_0=-0.9$ is more stable, the density threshold value here is higher. In fact, for $\xi_0=-0.5$, $(n_{EP}/n_e)_{thr}=0.025$, while for $\xi_0=-0.9$, $(n_{EP}/n_e)_{thr}=0.07$.

\subsection{Phase space analysis}
\label{secmpr}

\begin{figure}[H]
    \centering
    \includegraphics[height=7cm,width=10cm]{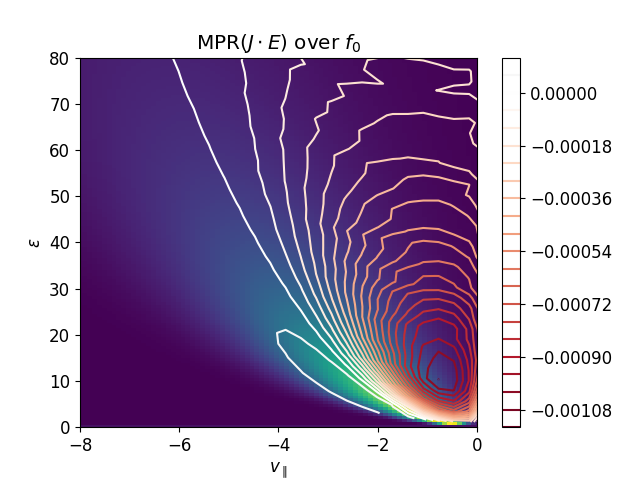}
    \caption{MPR diagnostic ($J\cdot E$) superimposed to $f_0(v_{\parallel
    },\varepsilon)$}
    \label{fig9.c}
\end{figure}

A mode-particle-resonance (MPR) diagnostic for ORB5 was developed in order to analyze which velocities of the energetic particle distribution were contributing the most to the excitation or damping of the mode \cite{novikau2019implementation}. The theoretical estimate for the resonance velocity comes from Eq.(41) of \cite{novikau2019implementation}:

\begin{equation}
    v_{\parallel, res}=qR_0\omega_{GAM} ,
    \label{eq2.d}
\end{equation}
where $\omega_{GAM}$ is the GAM frequency, $q$ the safety factor and $R_0$ the major radius. The reason for which we obtain such an expression of the resonance velocity will be outlined in section \ref{secanalysis} (see denominator of Eq. \ref{eq8.a}). As already said, for $\xi_0=-0.5,\;\sigma_{\xi}=0.2$, being $q\simeq2.3$, $R_0=1.66\,m$, $\omega_{GAM}=35560 rad/s$ the first resonant velocity is $v_{\parallel}=qR_0\omega_{GAM}=1.36\cdot10^5 m/s$. Normalized with respect to $v_s=\sqrt{k_BT_e/m_i}=1.88\cdot10^5 m/s$, we get $v_{\parallel,res,norm}=0.723$. If we plot the MPR diagnostic superimposed to the distribution function as in Fig.\ref{fig9.c}, we notice that the resonant velocity at which most of the power is exchanged between the particles and the mode is at $v_{\parallel,norm}=0.73$. This result shows the accordance between the theory and the numerical simulations. It is important to notice that the MPR is computed as the time derivative of the plasma kinetic energy
see Eq. (9) in \cite{novikau2019implementation}. This means that negative MPR contributions (the majority in Fig.\ref{fig9.c}) correspond to the plasma kinetic energy decrease due to the energy transfer to the field. Therefore it corresponds to a  positive power contribution to the mode, leading to the excitation of it. Therefore, in excited modes we'll see most of MPR plots to be rather negative and in damped ones to be positive. As we can see the exciting part of the distribution function has a positive gradient in $v_{\parallel}$, while we see also some portion of damping contribution on the left of the graph, in correspondence of negative gradients of $f_0$ with respect to $v_{\parallel}$. In general we can state:

\begin{align}
    & \frac{\partial f_0}{\partial |v_{\parallel}|}>0\implies J\cdot E <0\implies \text{excitation contribution}\\
    & \frac{\partial f_0}{\partial |v_{\parallel}|}<0\implies J\cdot E >0\implies \text{damping contribution}
\end{align}

\subsection{Effects of ion Temperature}

Changing the temperature of the thermal ions greatly affects their damping effect on the EGAM. Considering Eq. (\ref{eq2.d}), since GAMs can have higher poloidal sidebands with $m\geq1$, also the resonant velocity will have different values for the different sidebands:

\begin{equation}
    v_{\parallel, res}^m=\frac{qR_0\omega_{GAM}}{m}.
    \label{eq2.f}
\end{equation}

\begin{figure}[H]
    \includegraphics[width=6cm,height=5cm]{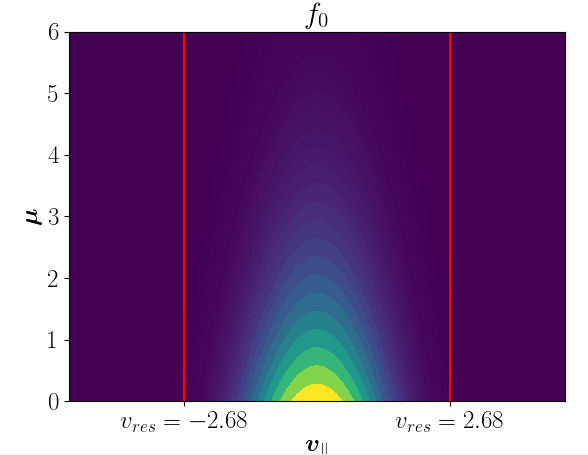}
    \includegraphics[width=6cm,height=5cm,tics=10]{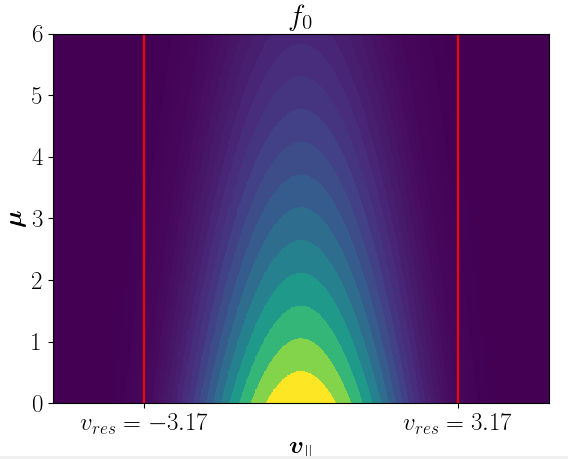}
    \caption[width=5cm,height=5.0cm]{Distribution functions of thermal ions in $(v_{\parallel},\mu)$ with $T_i/T_e=2.5$ (left) and $T_i/T_e=4.5$ (right), vertical lines highlight resonant velocities as computed from Eq. \ref{eq2.f}}
    \label{fig6.a}
\end{figure}

Nevertheless, numerical computations showed that the energy transfer between mode and thermal particles occurs mostly at the first resonant velocity \cite{novikau2019implementation}. The gaussian widt of the distribution function of thermal ions is proportional to the square root of temperature as shown in Fig.\ref{fig6.a}.

\begin{figure}[H]
    \centering
    \includegraphics[height=5.5cm]{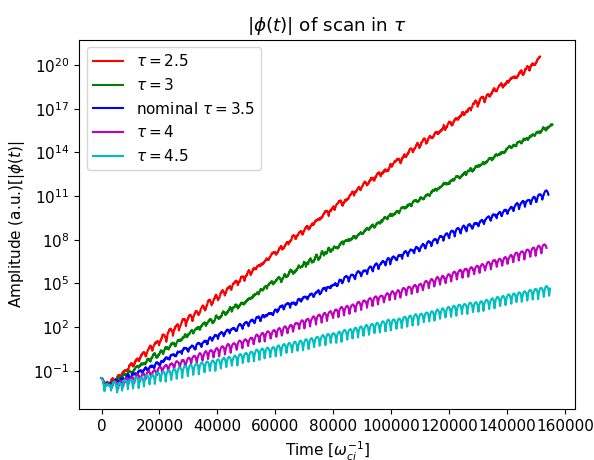}
    \caption{Mode amplitudes of a scan in $\tau$, ranging from 2.5 to 4.5, for NLED-AUG case}
    \label{fig7.a}
\end{figure}

\begin{figure}[H]
    \centering
    \begin{overpic}[width=10cm,tics=10]{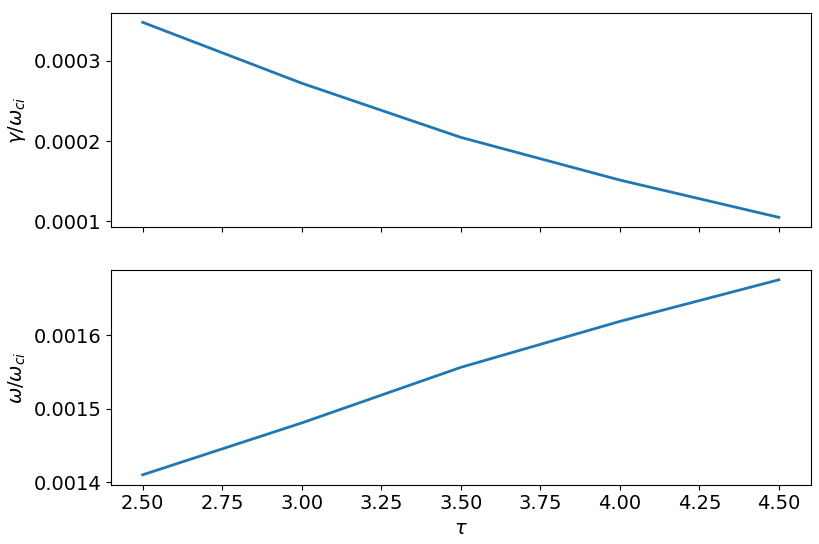}
        \put (-6,51) {\color{black}\Large$\displaystyle a)$}
        \put (-6,18) {\color{black}\Large$\displaystyle b)$}
    \end{overpic}
    \caption{Growth rate $\gamma/\omega_{ci}$ (a) and frequency $\omega/\omega_{ci}$ (b) of the scan in $\tau$ shown in Fig.\ref{fig7.a}}
    \label{fig8.a}
\end{figure}

As we see in the Figure \ref{fig6.a} the $v_{\parallel,res}$ intersect the ion distribution function in denser areas of the distribution function for higher values of $\tau=\frac{T_i}{T_e}$, and viceversa for lower values of $\tau$. This means that the mode is able to redistribute more energy to the thermal ion species if its temperature is higher leading to a stronger damping effect by the thermal species. It is interesting to notice that the resonant velocities for higher temperatures are higher, this is due to the fact that less excited modes (those obtained for higher $\tau$) have higher frequencies (close to $\omega_{GAM}$). In order to analyze the effects of $\tau$ of the ions, a scan in $\tau$ has been performed by varying $T_i$ in a range of values using the configuration on NLED-AUG case with $\xi_0=-0.5$ and $\sigma_{\xi}=0.3$. Figure \ref{fig7.a} and \ref{fig8.a} show clearly the effect of ion temperature. In Fig.\ref{fig7.a} the modes are growing faster for lower values of $\tau$. In figure \ref{fig8.a} the growth rates are plotted as function of $\tau$ and we can see the trend is monotonically decreasing. Meanwhile, the frequency has the opposite behaviour increasing with $\tau$. Once more this is due to the transition from an excited and undamped low frequency EGAM to a undamped high frequency GAM. Furthermore the GAM frequency can be written as \cite{qiu2010nonlocal}:

\begin{equation}
    \omega_{GAM}=\frac{v_{th}}{R_0}\sqrt{\left(\frac{7}{4}+\frac{T_e}{T_i}\right)}\;,
\end{equation}
where $v_{th}=\sqrt{2T_i/m_i}$. Hence, for higher values of $T_i$ correspond higher values of GAM frequency.

\section{Linear dispersion relation}
\label{secanalysis}

In this section we offer an analytical explanation to the effects shown in Section \ref{secorb5}. 
The derivation of the dispersion relation follows closely the steps presented in Ref. \cite{qiu2010nonlocal,chavdarovski2021linear}. The perturbed distribution function $\delta f$ can be represented in the following form:

\begin{equation}
    \delta f=Q\frac{\partial f_0}{\partial \varepsilon}\frac{\delta \phi}{m}+\exp{i\frac{mc}{
    QB
    ^2}\textbf{k}\times\textbf{
    B}\cdot\textbf{v}}\delta 
    H_g,
\end{equation}
where $\delta f$ has been split into an adiabatic response depending on the perturbed scalar potential $\delta \phi$ and a non-adiabatic part $\delta H_g$. The latter satisfies the linear gyrokinetic equation \cite{chen1987waves}:

\begin{equation}
    \left(\omega-\omega_d+i\omega_{tr}\frac{\partial}{\partial \theta}\right)\delta H_g=-\frac{Q}{m}\frac{\partial f_0}{\partial \varepsilon}J_0(k_{\bot}\rho_L)\omega\delta\phi.
    \label{eq3.a}
\end{equation}
The transit frequency $\omega_{tr}=v_{\parallel}/qR_0$, the particle drift frequency is $\omega_d=\overset{\wedge}{\omega}_dsin\theta=-k_r(v_{\bot}^2+2v_{\parallel}^2)/(2\Omega R_0)$, $\theta$ is the poloidal angle coordinate, $k_{\bot}$ is the perpendicular wave number, for GAMs $k_{\bot}\sim k_r$, $\rho_L=mcv_{\bot}/QB$ is the Larmor radius. $\Omega=QB/mc$ is the gyrofrequency, $J_0$ is the first kind Bessel function accounting for the finite Larmor radius (FLR) effects, $Q$ is the particle charge and the energy per unit mass of the particle is $\varepsilon=(v_{\parallel}^2+v_{\bot}^2)/2$.
Considering adiabatic electrons ($\omega/\omega_{tr,e}\sim\sqrt{m_e/m_i}\ll1$) and ignoring the FLR effects of electrons, Eq. \ref{eq3.a} can be trivially solved for electrons and the quasi-neutrality condition can be written as \cite{qiu2010nonlocal}:

\begin{equation}
    \frac{e}{T_e}(n_e+n_h)(\delta \phi-\overline{\delta \phi})=-\frac{e}{T_e}n_c\delta\phi+\langle J_0(k_{\bot}\rho_{L,c})\delta H_{g,c}\rangle+\langle\frac{e}{m}\frac{\partial f_{0,h}}{\partial \varepsilon}\delta\phi+J_0(k_{\bot}\rho_{L,h})\delta H_{g,h}\rangle,
    \label{eq4.a}
\end{equation}
where the bar $\overline{(...)}$ represents a magnetic surface averaged quantity and the operator $\langle...\rangle=\int...dv^3$ a velocity space integration, the subscripts $c,h$ refer to thermal (cold) and energetic (hot) ion species, respectively. 

We adopt $\delta=n_h/n_c\ll1$ as a smallness parameter and assume $T_c/T_h=O(\delta)$ keeping $\beta_h/\beta_c\sim1$, with $\beta$ being pressure to magnetic pressure ratio. In order to maximize the resonance drive for fast particles, we assume $\omega\sim\omega_{tr,h}$, $\omega_{d,c}/\omega\sim k_r\rho_{d,h}\sim O(\delta^{1/2})$ and $k_r\rho_{L,h}\sim O(\delta)$, with $\rho_{d,h}$ radial drift. For the thermal ion species: the radial drift is  $k_r\rho_{d,c}\sim O(\delta)$ and the Larmor radius effects are of higher order: $k_r\rho_{L,c}\sim O(\delta^{3/2})$. We can then expand the perturbed potential and non-adiabatic response as a power series of $\delta^{1/2}$: $\delta \phi=\overline{\delta \phi}+\widetilde{\delta\phi}^{(1/2)}+\widetilde{\delta\phi}^{(1)}+\widetilde{\delta\phi}^{(3/2)}+...$ and $\delta H_g=\overline{\delta H_g}+\widetilde{\delta H_g}^{(1/2)}+\widetilde{\delta H_g}^{(1)}+\widetilde{\delta H_g}^{(3/2)}+...$. Exploiting such expansion, Eq. (\ref{eq3.a}) can be rewritten for all the orders as in Eq.(3-9) of \cite{qiu2010nonlocal}. Such system can be combined with quasi-neutrality Eq. (\ref{eq4.a}) and solved order by order (up to the 3$^{rd}$ order) we obtain the EGAM dispersion relation (see \cite{qiu2010nonlocal} for details):

\begin{equation}
    -1+\frac{\omega_G^2}{\omega^2}+\frac{\overline{\delta n_h}}{\overline{\delta \phi}}\frac{m\Omega}{en_ek_r^2}=0,
    \label{eq5.a}
\end{equation}
where $\omega_G=\left(\sqrt{\frac{7}{4}+\frac{T_e}{T_i}}\right)\frac{v_{th}}{R_0}$ is the GAM frequency and $v_{s,i}$ is the core thermal velocity. The last term on the left hand side of Eq. (\ref{eq5.a}) contains the integral in velocity space of the non-adiabatic component of the perturbation of the distribution function, written in phase-space coordinates ($\varepsilon,\;\xi$) as:

\begin{equation}
    \overline{\delta n_h}=\int\overline{\delta H_h^{(3)}}dv^3=2\pi\int_{-1}^1\int_0^{\infty}\overline{\delta H_h^{(3)}}(-\sqrt{2\varepsilon})d\varepsilon d\xi.
    \label{eq6.a}
\end{equation}

For passing particles, $\overline{\delta H_h^{(3)}}=-\frac{\overset{\wedge}{\omega}_d^2\left(\frac{e}{m}\frac{\partial f_0}{\partial \varepsilon}\right)}{2(\omega^2-\omega_{tr}^2)}\overline{\delta\phi}$. Being $\frac{\partial f_0}{\partial \varepsilon}=\frac{\partial f_0}{\partial \varepsilon}+\frac{\partial \xi}{\partial \varepsilon}\frac{\partial f_0}{\partial \xi}$, applying it to the slowing down with pitch dependency previously defined (Eq. (\ref{Eq1.a})) the full derivative is:
\begin{equation}
    \frac{\partial f_0}{\partial\varepsilon}=f_0(\varepsilon,\xi)\left(-\delta(\varepsilon_{\alpha}-\varepsilon)-\frac{3}{2}\frac{\varepsilon^{\frac{1}{2}}}{{\varepsilon^{3/2}+\varepsilon_c^{3/2}}}+\frac{\xi}{\varepsilon}\frac{(\xi-\xi_0)}{2\sigma_{\xi}^2}\right)
    \;,
    \label{eqcorr11}
\end{equation}
and the integral in Eq. (\ref{eqcorr11}) can be rewritten explicitly in terms of our coordinates as:

\begin{equation}
\begin{aligned}
    &\overline{\delta n_h}  =2\pi\frac{k_r^2\overline{\delta\phi}q^2e}{4\Omega^2m}\int_{-1}^1\int_0^{\infty}\sqrt{2\varepsilon}\frac{(\varepsilon(1+\xi^2))^2}{q^2R_0^2\omega^2-2\varepsilon\xi^2}\frac{\partial f_0}{\partial \varepsilon} d\varepsilon d\xi =\\
    = & 2\pi K\int_{-1}^1\int_0^{\infty}\frac{\varepsilon^{\frac{5}{2}}(1+\xi^2)^2}{q^2R_0^2\omega^2-2\varepsilon\xi^2} exp\left(-\frac{(\xi-\xi_0)^2}{2\sigma_{\xi}^2}\right)\frac{\Theta(\varepsilon_{\alpha}-\varepsilon)}{\varepsilon^{3/2}+\varepsilon_c^{3/2}}\\
    &\left(-\delta(\varepsilon_{\alpha}-\varepsilon)-\frac{3}{2}\frac{\varepsilon^{\frac{1}{2}}}{{\varepsilon^{3/2}+\varepsilon_c^{3/2}}}+\frac{\xi}{\varepsilon}\frac{(\xi-\xi_0)}{2\sigma_{\xi}^2}\right)d\varepsilon d\xi,
    \label{eq8.a}
\end{aligned}
\end{equation}
where
\begin{equation}
    K=\sqrt{2}\frac{2\sqrt{\frac{2}{\pi}}}{\sigma_{\xi}[erf(\frac{\xi_0+1}{\sqrt{2}\sigma_{\xi}})-erf(\frac{\xi_0-1}{\sqrt{2}\sigma_{\xi}})]}\frac{0.75\; n_{val}}{\pi }ln(1+(\frac{\varepsilon_{\alpha}}{\varepsilon_c})^{3/2})\frac{k_r^2\overline{\delta\phi}q^2e}{4\Omega^2m}
\end{equation}

Explicit integration in velocity space as in \cite{qiu2010nonlocal} is not possible since the distribution function is not characterized by a Dirac function. Not even an explicit integration as in \cite{girardo2014relation} is possible, because the characteristic structure of the plasma dispersion function  ($Z(z)=\frac{1}{\sqrt{\pi}}\int_{-\infty}^{\infty}\frac{exp(-y^2)}{y-z}dy$) is not present in equation (\ref{eq8.a}).
Nevertheless, it's possible to study the stability of the EGAM, noticing that the imaginary part of $\omega$ (namely the growth rate of the mode) in equation (\ref{eq5.a}) is positive if the imaginary part of $\overline{\delta n_h}$ is positive \cite{chavdarovski2021linear}. Without integrating eq. (\ref{eq8.a}), it's possible to plot the integrand function for an interval of energies and parallel velocities, imposing the $\omega$ inside eq. (\ref{eq8.a}) to equal its value computed from the simulations (see Section \ref{secorb5}). Changing the parameters $\xi_0$ and $\sigma_{\xi}$, we can see that the integrand's sign is rather positive at most of the energy levels as we pick values of $\xi_0$ far from the extremes 0 and -1, and low values of $\sigma_{\xi}$. This result is consistent with the numerical results obtained in Section \ref{secorb5}. 

Some plots of the integrand computed at $\varepsilon\sim30keV$ as function of $\xi$ are shown below in Fig.\ref{fig2.b}. The only growing case of the three shown is the ($\xi_0=-0.5,\;\sigma_{\xi}=0.2$) one (Fig.\ref{fig2.b}.a), the other ones are damped. As it was already discussed, this is due to the fact that when $\sigma_{\xi}\longrightarrow\infty$ the distribution function becomes isotropic in phase space and, hence, unable to excite n=0 modes (Fig.\ref{fig2.b}.b), (as mentioned in the Introduction, n=0 modes need $f_0$ gradients in velocity space to be excited \cite{betti1992stability}). Meanwhile, too low values of $\xi_0$ correspond to the case where most of the particles are trapped, mitigating the effects of inverse Landau damping and hindering the excitation of the mode (Fig.\ref{fig2.b}.c). For values of $\xi_0$ close to -1, we have a specular effect, most of the energetic particles are deeply passing and hence are unable to exchange energy with the mode at $v_{\parallel,res}$, hindering again the excitation of the mode.

\begin{figure}[H]

\centering
\includegraphics[width=.33\textwidth]{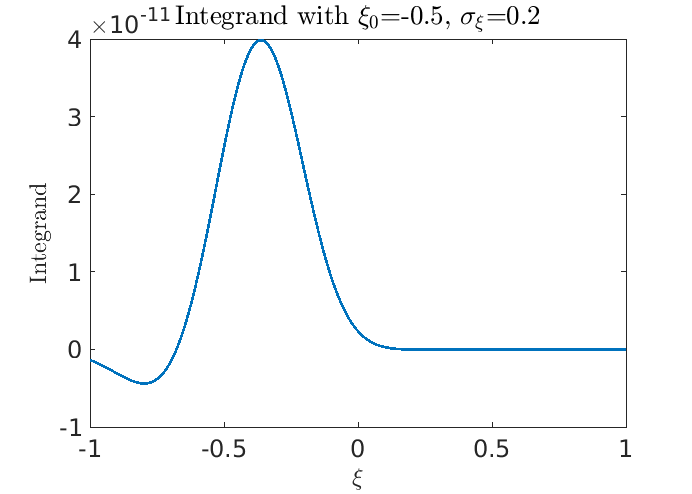}\hfill
\includegraphics[width=.33\textwidth]{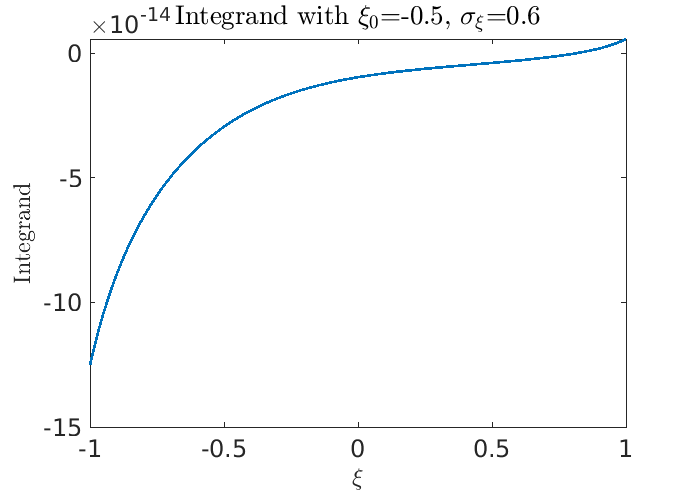}\hfill
\includegraphics[width=.33\textwidth]{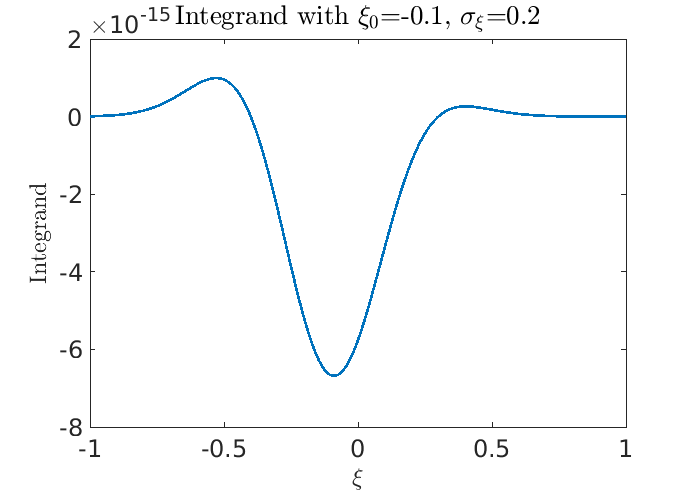}

\caption{Plot of the integrand of Eq. (\ref{eq8.a}) at $\varepsilon\sim30keV$ for $\xi_0=-0.5,\;\sigma_{\xi}=0.2$ (a);$\xi_0=-0.5,\;\sigma_{\xi}=0.6$ (b), $\xi_0=-0.1,\;\sigma_{\xi}=0.2$ (c)}
\label{fig2.b}

\end{figure}

This method is useful to theoretically validate the numerical results for extreme cases (strongly excited/damped modes), however it's very hard to use it for an evaluation of an excitation threshold limit in terms of $n_{EP}/n_e,\;\xi_0 \text{ or } \sigma_{\xi}$. In fact, for not strongly excited or damped modes, the integrand is partially negative and positive for different values of $\xi$ and $\varepsilon$. In order to evaluate threshold values we have to consider the results obtained from ORB5 simulations presented in Section \ref{secorb5}.

These kinds of plots are also useful to evaluate the main resonance velocity. Setting the growth rate to 0, the denominator of the term $\frac{\varepsilon^{\frac{5}{2}}(1+\xi^2)^2}{q^2R_0^2\omega^2-2\varepsilon\xi^2}$ will go to 0 marking a resonant condition for the power exchange between particles and the mode. For example, for $\xi_0=-0.5,\;\sigma_{\xi}=0.2$, the first resonant velocity is $v_{\parallel}=\sqrt{2\varepsilon\xi^2}=\sqrt{q^2R_0^2\omega^2}=1.36\cdot10^5 \frac{m}{s}$. This value is comparable to the result from the Mode-Particle-Resonance (MPR) diagnostic \cite{novikau2019implementation}, obtained from ORB5 simulations. 

\subsection{Simplified dispersion relation}

Simplified dispersion relation can be obtained if we consider the distribution in $\xi$ to be close to
$\delta(\xi-\xi_0)$, in which case we have $f_0=n_h/(2\sqrt{2}\pi\, ln(\varepsilon_c/\varepsilon_\alpha)) \delta(\xi-\xi_0)\Theta(1-\varepsilon/\varepsilon_\alpha) $. Following the derivations envisaged in Refs. \cite{qiu2010nonlocal,chavdarovski2021linear}
we obtain
\begin{eqnarray}\label{e:disper}
\hskip -4em -1 +\frac{\omega_G^2}{\omega^2}+ N_b \left[\frac{3}{2}(1-\xi_0^2+4 \xi_0^4)\ln \left(1-\frac{\omega_{ts}^2}{\omega^2}\right) +
\left( \frac{1}{1-\omega_{ts}^2/\omega^2}-1\right)\right]=0 \,.
\end{eqnarray}
Here, $\omega_{ts}= \sqrt{2 \varepsilon_\alpha}\,\xi_0 / (q R_0) $ is the EP transit frequency and $N_b=q^2 /(4\xi_0^2 \ln (\varepsilon_\alpha/\varepsilon_c))\, n_h/n_c$. The GAM frequency with Landau damping due to thermal ions $\omega_{G}$ can be found
as a solution of the equation (from  \cite{zonca2008radial}):
$$\omega=-q^2\omega_{Ti}\left(F(\omega/\omega_{Ti})+N^2(\omega/\omega_{Ti})/D(\omega/\omega_{Ti})\right)\,,$$
with $F (x) = x(x^2 + 3/2) + (x^4 + x^2 + 1/2)Z(x)$, $N(x) =
x + (1/2 + x^2 )Z(x)$ and $D(x) = Z(x) + (1 + T_i /T_e )(1/x)$ \cite{zonca1996kinetic}.
Eq.~(\ref{e:disper}) is strictly valid for well circulating ions, i.e. $\xi_0 \sim \pm 1$.
The logarithmic term in this equation, for $\omega<\omega_{ts}$ has complex values and gives a drive if the constant in-front of it is positive, which is always the case here. This shows there is no threshold in $\xi_0$ related to the distribution function,
however some drive is necessary to overcome the finite Landau damping. The last term in Eq.~(\ref{e:disper}), along with the real part of the logarithmic term is responsible for shifting the frequency of the mode below the GAM frequency. Eq.~(\ref{e:disper}) can be easily solved numerically to obtain the modes frequency and growth rate. A rough estimate of the growth rate can be given in the limit $\omega\gg \omega_{Ti}$ and $\gamma/\omega_r\ll1$, where the GAM frequency can be approximated with
$$\omega^2=q^2\omega_{Ti}^2\left(\frac{7}{4}+\frac{T_e}{T_i}\right)- i \sqrt{\pi}q^2\frac{\omega^5}{\omega_{Ti}^3} e^{-\omega^2/\omega_{Ti}^2}\,,$$
giving a growth rate of
\begin{eqnarray}\label{e:gamma}
\hskip -4em \gamma/\omega_r \approx q^2 \left( \frac{3 \pi}{16 \ln (E_b/E_c)\xi_0^2}\,\frac{n_h}{n_c}(1-\xi_0^2+4 \xi_0^4)-
\frac{\sqrt{\pi}}{2}\frac{\omega_r^3}{\omega_{Ti}^3}e^{-\omega_r^2/\omega_{Ti}^2}\right)\,\,,
\end{eqnarray}
with $\omega_r$ being real frequency of the mode. This equation shows a threshold in the density ratio $n_h/n_c$, as well as dependence of the growth rate on the safety factor \cite{qiu2010nonlocal,chavdarovski2021linear}. The growth rate given here is the upper limit of this distribution with maximal anisotropy.

\section{RABBIT distribution functions}
\label{secrabbit}

\begin{figure}[H]
    \centering
    \includegraphics[height=8cm]{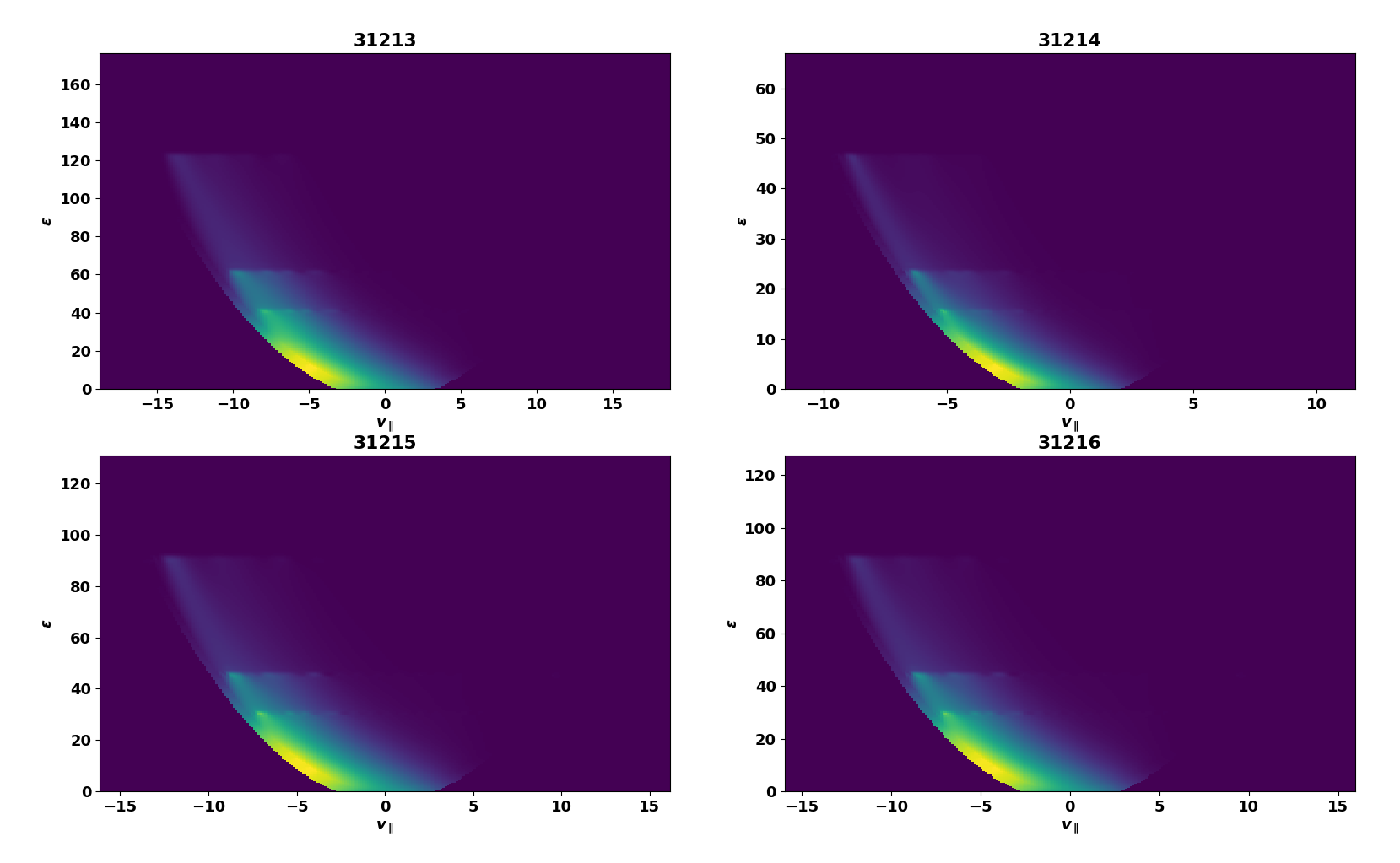}
    \caption{Realistic distribution functions generated from RABBIT, NLED-AUG case shots $\#$ 31213-4-5-6}
    \label{fig9.a}
\end{figure}

RABBIT has been run to reproduce the fast particle distribution functions generated by NBI in NLED-AUG discharges \#31213-6 \cite{lauber2014off}. The four different distribution functions are plotted in Figure \ref{fig9.a}. They show a strong anisotropy in $v_{\parallel}$, this is the reason that led us to build an analytical distribution function such as described in Section \ref{secf0}. Furthermore, the distribution functions consist of three different injection velocities $E,\;E/2,\;E/3$, as typical for positive NBI injectors. (The injection velocities in the figure seem to be different from shot to shot because of the normalization with respect to the electron temperature, on an absolute scale all the injection velocities are the same). These distribution functions originated from three different NBIs angles with respect to the axis of the machine. Shot 31213 has the most off-axis angle (7.13$^{\circ}$), shot 31214 the most on-axis (6.07$^{\circ}$) and 31215-6 a mid-range angle (6.65$^{\circ}$).

\begin{figure}[H]
    \centering
    \includegraphics[height=8cm]{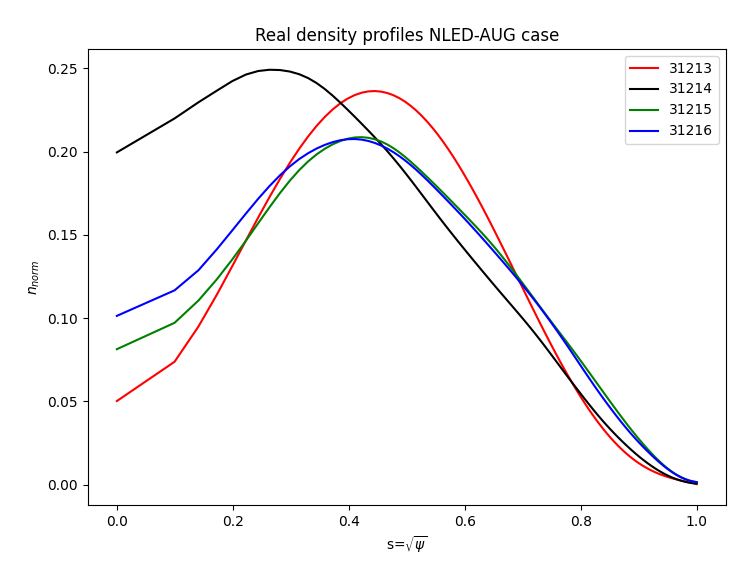}
    \caption{Realistic EP density profiles, NLED-AUG case shots $\#$ 31213-4-5-6, as function of $s=\sqrt{\psi}$ with $\psi$ magnetic flux coordinate}
    \label{fig15.a}
\end{figure}

In order to achieve experimental relevant scenarios, these distribution functions were given as input to the gyrokinetic code ORB5 with the realistic profiles of density and temperature from the experimental setup of the different shots. The EP realistic profiles from the four shots are represented in Figure \ref{fig15.a}

\begin{figure}[H]
    \includegraphics[width=11cm,height=6cm]{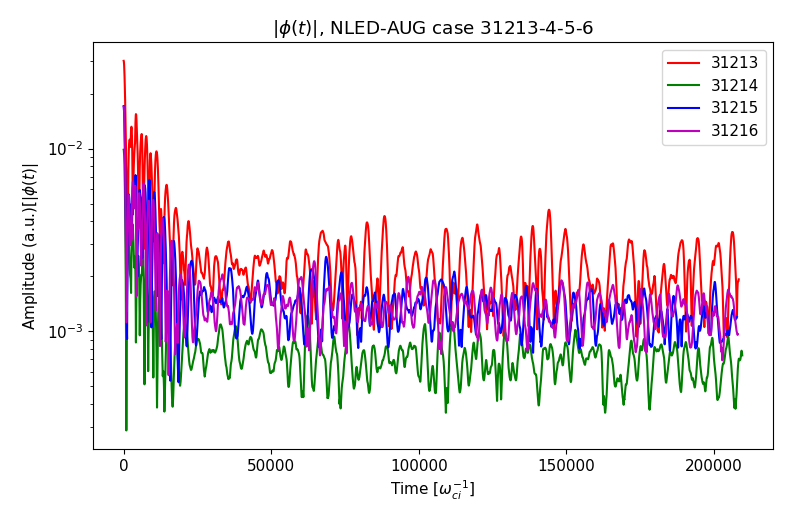}
    \caption[width=5cm,height=5.0cm]{NLED-AUG case modes for the four shots, using the four different distribution functions showed in Fig.\ref{fig9.a}}
    \label{fig16.a}
\end{figure}

The results obtained from ORB5 using the RABBIT distribution functions are shown in Fig.\ref{fig16.a}. As we notice, the growth rates are  all negative. In fact, such NBI angles correspond to high $\xi_0$ values and, as we have seen from Section \ref{secorb5}, such high values of $\xi_0$ produce rather stable modes.

\subsection{MPR diagnostic from RABBIT distribution functions}

\begin{figure}[H]
    \centering
    \includegraphics[width=12
cm,height=8cm]{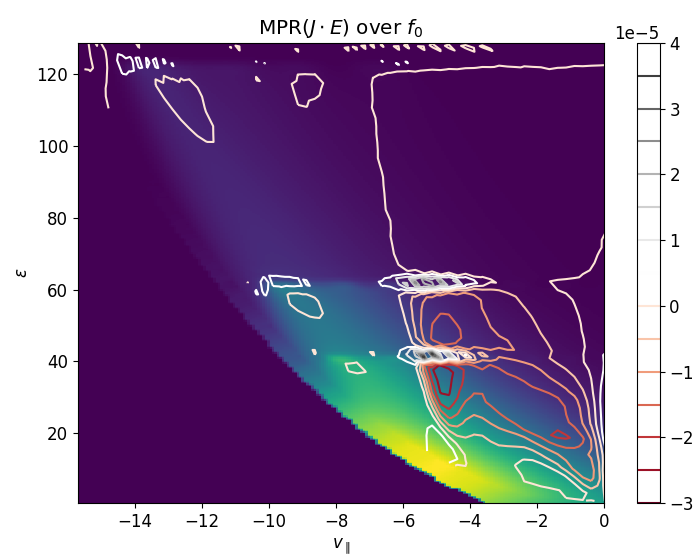}
    \caption{MPR diagnostic ($J\cdot E$) superimposed to $f_0(v_{\parallel},\varepsilon)$ of NLED-AUG case shot $\#$ 31213}
    \label{fig17.a}
\end{figure}

We repeat here the considerations previously done in Section \ref{secorb5}, presenting the MPR diagnostic results for the simulation of shot 31213. We show a graph similar to that shown in Figure \ref{fig9.c}, this time it represents the MPR diagnostic contour plot superimposed to the RABBIT distribution function for the shot 31213 (Fig.\ref{fig17.a}).
Again we compute the theoretical estimate for the parallel resonant velocity, being $q\simeq2.3$, $R_0=1.66$ m and $\omega_{GAM}=2.04\cdot10^5$ rad/s:

\begin{equation}
    \frac{v_{\parallel,res}}{v_{th,i}}=\frac{qR_0\omega_{GAM}}{v_{th,i}}=4.15.
    \label{eq14.a}
\end{equation}

Figure \ref{fig17.a} offers a much more complex image than Fig.\ref{fig9.c}. Here there are different peaks, both negative and positive, positioned in different points of the distribution function. In the plot we can see that the highest negative peak, at which the power exchange takes place, is located at $v_{\parallel}\simeq-4.75$. This value is very close the main resonant velocity found in Eq. (\ref{eq14.a}). Furthermore, according to Eq. (42) of \cite{novikau2019implementation}, there are many resonant velocities due to the exchange of energy taking place at higher poloidal harmonics ($|m|\geq1$). Therefore, we can find other resonant velocities as:

\begin{equation}
    v_{\parallel, res}^{(m)}=\frac{qR_0\omega_{GAM}}{m}.
\end{equation}

In fact, we notice a smaller peak indeed positioned in correspondence of another resonant velocity at $v_{\parallel}=-1.34$ (Fig.\ref{fig17.a}). Considering m=3, and considering the result from Eq.(\ref{eq14.a}), $v_{\parallel, res}^{(3)}\simeq1.38$, which is very close to the secondary resonant velocity at which some of the power is exchanged in Fig.\ref{fig17.a}.
In the figure we see also other negative peaks either near the main resonance velocity or the other poloidal mode velocities, at higher energies. It is interesting to notice the position of the positive (damping) peaks too. Most of them in fact are disposed just above each of the three injection velocities. This result is somehow in accordance with the theory \cite{betti1992stability}, in fact, the parts of the distribution function where $\frac{\partial f_0}{\partial\varepsilon}<0$ are those damping the mode. In correspondence of the injection velocities we have very steep negative gradients which damp the mode. (In fact, the largest part of particles in these discontinuities can be only accelerated generating a damping effect on the mode.) 

\subsection{Results and threshold values for RABBIT distribution functions}

In order to qualitatively compare the numerical results from ORB5 with the experimental data (Sec. \ref{subsecexperim}), it is of interest to study the threshold values for the RABBIT distribution functions too. To this purpose, a scan in the fraction of EPs has been run as in Section \ref{subsecfpartorb5}. We have therefore set $n_{EP}/n_e$ to different values comprehended between 2$\%$ and 50$\%$. The results have been plotted for the different distribution functions. We analyze case by case the three shots 3121{3,4,6}.

\subsubsection{EP density threshold values for shot 31213}

\begin{figure}[H]
    \includegraphics[width=6cm,height=5cm]{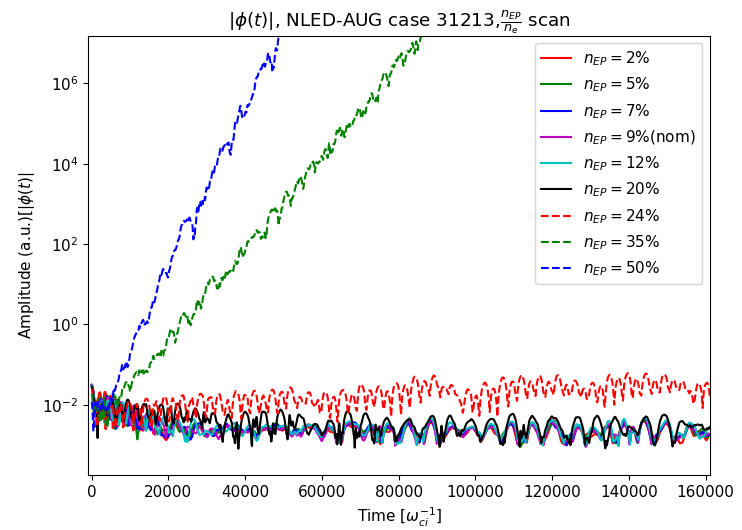}
    \begin{overpic}[width=6cm,height=5cm,tics=10]{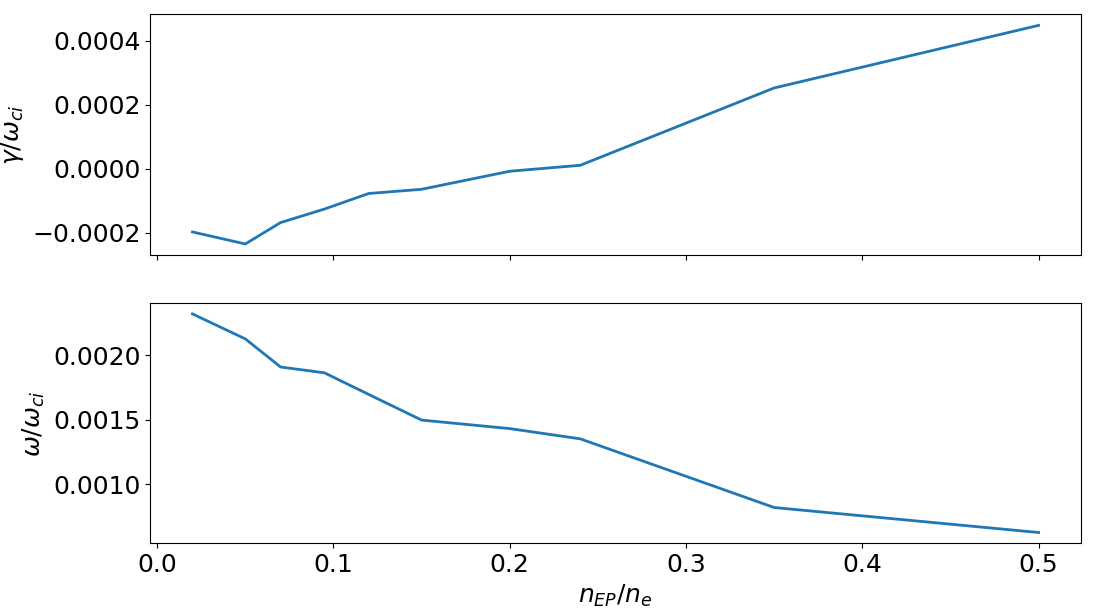}
        \put (-2,73) {\color{black}\large$\displaystyle a)$}
        \put (-2,33) {\color{black}\large$\displaystyle b)$}
    \end{overpic}
    \caption[width=5cm,height=5.0cm]{$\frac{n_{EP}}{n_e}$ scan for NLED-AUG case shot 31213, modes in time (left), modes' growth rates (a) and frequencies (b) as function of $\frac{n_{EP}}{n_e}$ (right)}
    \label{fig18.a}
\end{figure}

In this scan we ran the NLED-AUG case with ORB5 with the distribution function and the profiles from shot 31213. The scan was performed up to an EP concentration $\sim50\%$. Such a high concentration represents a sort of upper limit in this kind of scans. In fact, being $n_{EP}/n_e$ a volume averaged quantity for values higher than this we seriously risk to reach local values of EP concentration higher than the electron one, leading to non-physical results in the ORB5 code. Furthermore, in such extreme cases all the quantities depending on bulk plasma parameters lose their meaning, for example the GAM frequency, which depends on $T_i$. 
In Fig.\ref{fig18.a} we can see the different modes in time on the left, and the modes' growth rates and frequencies as function of $n_{EP}/n_e$ on the right. We notice the continuous increase of growth rate with the fraction of EP. On the other hand, as expected, frequencies decrease as the modes are transitioning from damped GAMs to excited EGAMS. As we can see from the plots, we could fix a threshold value for case 31213 at $(n_{EP}/n_e)_{thr}=0.23$.

\subsubsection{EP density threshold values for shot 31214}

\begin{figure}[H]
    \includegraphics[width=6cm,height=5cm]{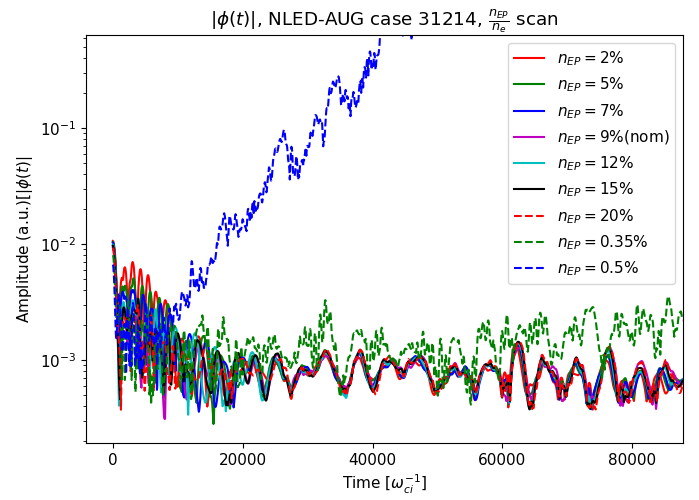}
    \begin{overpic}[width=6cm,height=5cm,tics=10]{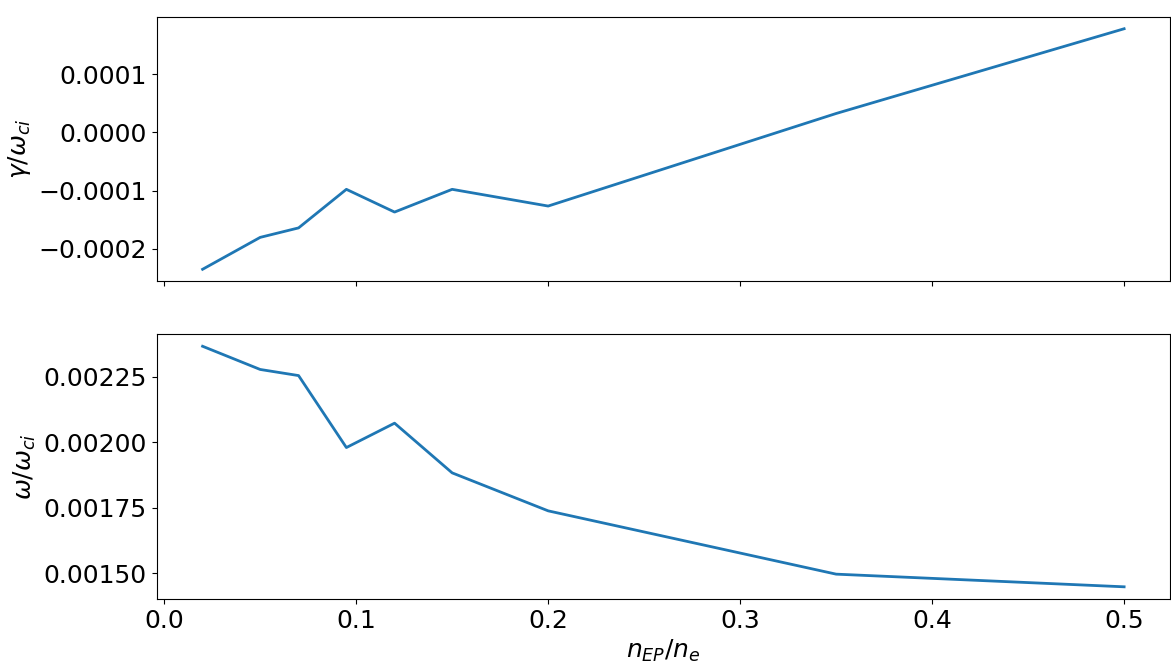}
        \put (-2,73) {\color{black}\large$\displaystyle a)$}
        \put (-2,33) {\color{black}\large$\displaystyle b)$}
    \end{overpic}
    \caption[width=5cm,height=5.0cm]{$\frac{n_{EP}}{n_e}$ scan for NLED-AUG case shot 31214, modes in time (left), modes' growth rates (a) and frequencies (b) as function of $\frac{n_{EP}}{n_e}$ (right)}
    \label{fig19.a}
\end{figure}

We repeat the same scan for NLED-AUG case $\#$ 31214. This time we also extend our scan $n_{EP}/n_e=0.5$, for the excitation of the mode can be found just for higher EP fractions. Fig.\ref{fig19.a} is structured as Fig.\ref{fig18.a}. Here, too, we find the usual behaviour of growth rate and frequency, as one increases while the other decreases as we increase the fraction of energetic particles. This time, the threshold value is found to be higher than in case 31213 (Fig.\ref{fig19.a}), being more stable and the beam more on-axis than case 31213 case, the threshold value is higher, as already noticed in Section \ref{subsecfpartorb5}. As we can see from the plots, we could fix a threshold value for case 31214 at $(n_{EP}/n_e)_{thr}\simeq0.32$.

\subsubsection{EP density threshold values for shot 31216}

\begin{figure}[H]
    \includegraphics[width=6cm,height=5cm]{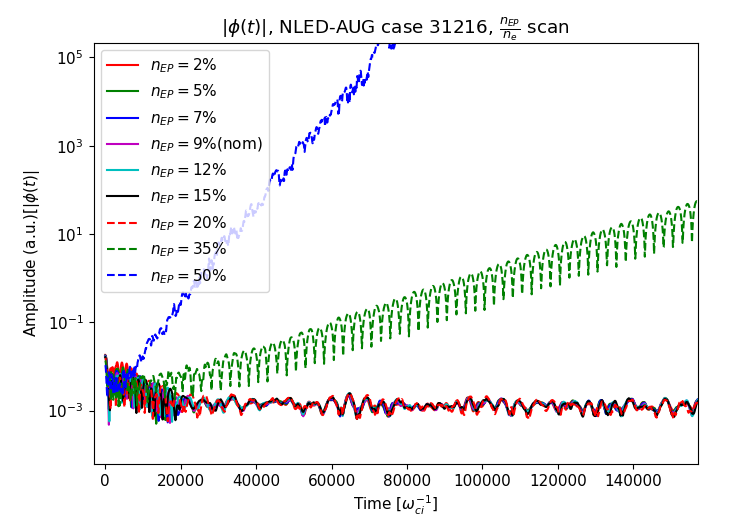}
    \begin{overpic}[width=6cm,height=5cm,tics=10]{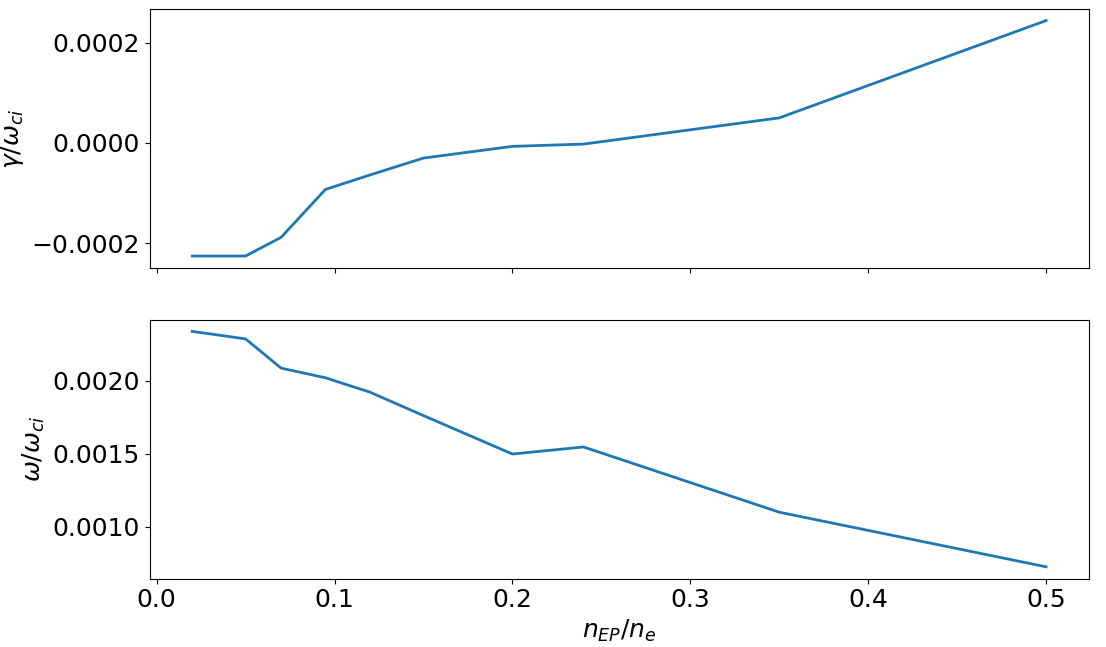}
        \put (-2,73) {\color{black}\large$\displaystyle a)$}
        \put (-2,33) {\color{black}\large$\displaystyle b)$}
    \end{overpic}
    \caption[width=5cm,height=5.0cm]{$\frac{n_{EP}}{n_e}$ scan for NLED-AUG case shot 31216, modes in time (left), modes' growth rates (a) and frequencies (b) as function of $n_{EP}/n_e$ (right)}
    \label{fig20.a}
\end{figure}

The density scan was repeated for case $\#$ 31216, as before we reached $(n_{EP}/n_e)=0.5$. Results are plotted in Fig.\ref{fig20.a}. The trend is the same as those found above: an increasing fraction of energetic particles implies an increase of growth rate and a reduction in frequency. Being the injection angle 6.65$^{\circ}$, we expect the growth rate threshold to be somewhere between the threshold values of shots 31213 and 31214. Actually, we find out that the result is as represented in Fig.\ref{fig20.a}, in this case $(n_{EP}/n_e)_{thr}\simeq0.27$. 

\subsection{Comparison with experimental measurements}
\label{subsecexperim}

\begin{figure}[H]
\hspace*{-2.5cm}
    \centering
    \includegraphics[height=8cm]{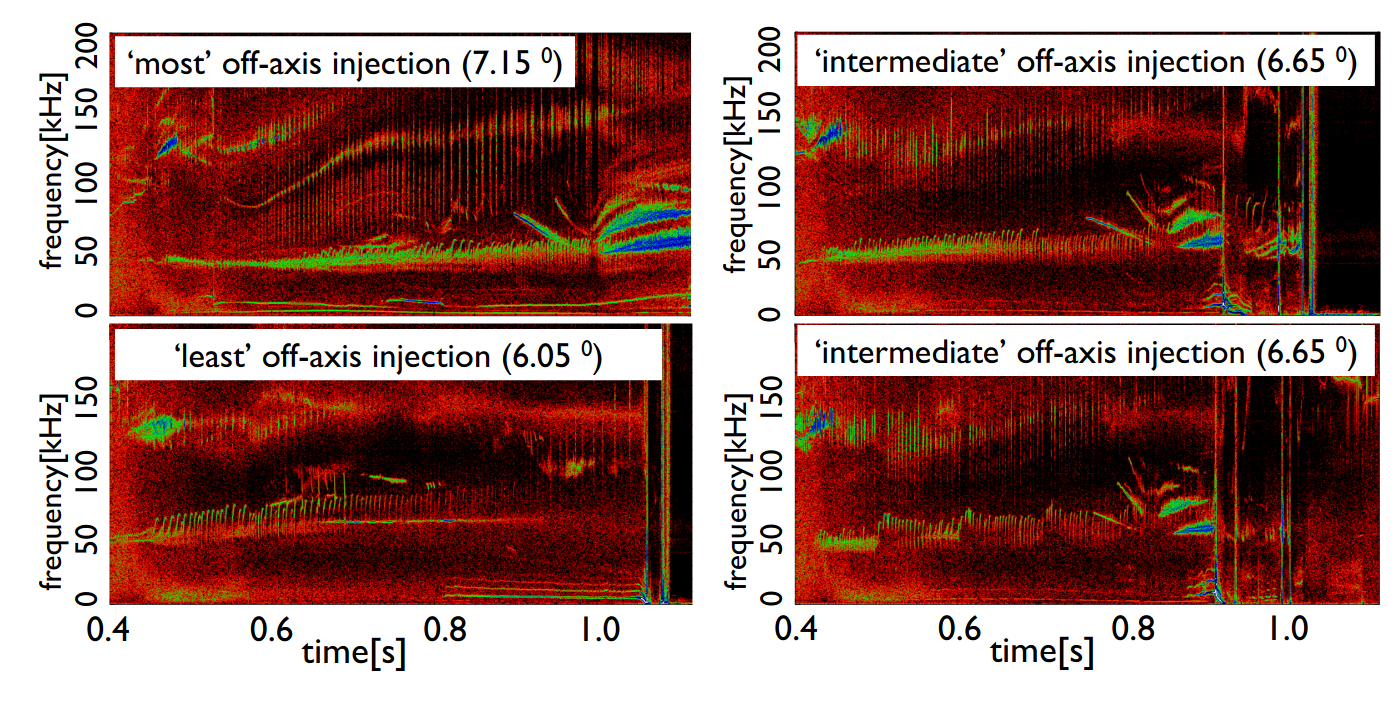}
    \caption{Experimental data from NLED-AUG case from magnetic pick-up coils (in order top-left, bottom-right: 31213,5,4,6)}
    \label{figexp}
\end{figure}

The results obtained in the previous section are very interesting and follow the theoretical and numerical expectations reported above. Nevertheless if we analyze the data obtained from the magnetic pick-up coils showing the structures of the magnetic perturbations in NLED-AUG cases in Fig.\ref{figexp}, we notice that the EGAM (green line at $\sim50 kHz$) is excited for every case. As expected, the most unstable case is 31213 and the least case 31214. In this respect the trends found analytically, numerically and experimentally are compatible. Nevertheless, we have seen that in numerical simulations, with an EP fraction of $(n_{EP}/n_e)\simeq9\%$, the modes are all damped. Threshold values for excitement in simulations with RABBIT $f_0$ are found at much higher EP fraction. This hints that there must be some non-linear effect, ignored in the electrostatic, linear simulations, that is driving EGAMs unstable even with such a small EP density. As theorized in \cite{qiu2016effects}, such effect could be due to non-linear interactions of n=1 Alfv\'enic modes and EGAMs. Reference \cite{qiu2016effects} proved that it is possible for Alfvén modes to trigger EGAMs. Eventually, it was also shown that such interactions work the other way round: EGAMs can non-linearly excite Alfvén waves \cite{vannini2021gyrokinetic}. Possibly, considering these non-linear interactions too, a better estimate of the growth rate will be obtained for the NLED-AUG experimental cases. This non-linear electromagnetic approach to the study of driving effects of EP on plasma instabilities will be addressed in future work.

\section{Conclusions}

In this paper we addressed the effects of anisotropic distribution functions on tokamak plasmas, and evaluated the effects of such distributions on the stability of Geodesic Acoustic Modes, taking in consideration as configuration the NLED-AUG case. Firstly, we implemented a new anistropic, asymmetric in $v_{\parallel}$ distribution function, parametrized through two variables : $\xi_0$ and $\sigma_{\xi}$ (section \ref{secf0}). Various scans have been run with the gyrokinetic code ORB5, and the threshold values have been obtained for the two parameters characterizing the distribution function: $\xi_0$ and $\sigma_{\xi}$, and for the density fraction $n_{EP}/n_e$(section \ref{secorb5}). It has been found that the modes gets unstable for low values of $\sigma_{\xi}$ and for values of $\xi_0$ included in an interval $(-0.9,-0.3)$. This interval widens for decreasing $\sigma_{\xi}$ and increasing EP fraction. The threshold value for $\frac{n_{EP}}{n_e}$ has been found to change according to the $\xi_0$, $\sigma_{\xi}$ configuration, yielding different values for different cases. The effects of ion temperature and the structure of the power exchange (MPR) have been verified too. The growth rate decreases as thermal ion temperature increases and it has been found that the theoretical main resonance velocity is actually the one at which most of the power is exchanged between the mode and the energetic particles.
A theoretical model has been proposed based on the derivation of the dispersion relation of EGAMs to demonstrate the stability conditions of the extreme cases for $\xi_0$ and $\sigma_{\xi}$. This method is not useful to evaluate threshold values, to that purpose the numerical simulations were used to derive such thresholds. Finally, experimental like distribution functions obtained from RABBIT have been used in the NLED-AUG cases 31213-6, with the experimental temperature and density profiles from the four different shots. 
Results showed all the modes were damped and the power exchange has been studied for the case 31213. A more complex structure with respect to the analytical distribution function was found this time. Interaction of EPs and the mode has been found to happen also at higher mode resonant velocities. Scans in EP fraction have been performed for the three cases 3121{3,4,6}, representing the 3 different injection angles of the NBIs. As expected different threshold values were found for the different injection angles. At the end, a qualitative comparison with experimental measurements from magnetic pick-up coils shows that at the standard EP fraction ($\sim9\%$) EGAMs were triggered in AUG in all the cases (even if with different growth rates). From the simulations with the experimental-like $f_0$ and profiles only the case 31213 is weakly unstable. This hinted that some kind of non-linear effect should be considered in order to get fully predictive simulations. Also some experimental uncertainties could contribute to this behaviour of the different shots. Firstly, only shot 31216 had $T_i$ measurements (through beam blips), the other $T_i$ profiles were actually estimated using the 31216$^{th}$ data. Actually, $T_i$ could be smaller in shots 31213/4/5, driving higher growth rates in the simulations. In second place, discharges have impurities, whose effects are neglected in ORB5 simulations. Finally, q profiles might have an error since the damping is $\sim e^{-q^2}$, shifting the stability boundary. We can assume the interactions between the n=1 Alfvén modes and the EGAMs could be driving EGAMs looking at the results from reference \cite{vannini2021gyrokinetic}. In fact, the anisotropic distribution function alone is not able to drive them linearly at such small EP fraction. This non-linear studies will be addressed in future work.

\section{Acknowledgment}

This work
has been carried out within the framework of the
EUROfusion Consortium and has received funding
from the Euratom research and training program 2014-
2018 and 2019-2020 under grant agreement No 633053.
The views and opinions expressed herein do not
necessarily reflect those of the European Commision.
Part of this work is supported by R\&D Program through the Korean Institute of Fusion Energy (KFE)
funded by the Ministry of Science and ICT of the Republic of Korea (KFE-EN2141-7).
Simulations presented in this work were performed on
the MARCONI FUSION HPC system at CINECA and
the HPC systems of the Max Planck Computing and
Data Facility (MPCDF).

\bibliographystyle{unsrt}
\bibliography{bibliography}

\begin{thebibliography}{10}

\bibitem{winsor1968geodesic}
Niels Winsor, John~L Johnson, and John~M Dawson.
\newblock Geodesic acoustic waves in hydromagnetic systems.
\newblock {\em The Physics of Fluids}, 11(11):2448--2450, 1968.

\bibitem{smolyakov2016dispersion}
AI~Smolyakov, MF~Bashir, AG~Elfimov, M~Yagi, and N~Miyato.
\newblock On the dispersion of geodesic acoustic modes.
\newblock {\em Plasma Physics Reports}, 42(5):407--417, 2016.

\bibitem{zonca1996kinetic}
Fulvio Zonca, Liu Chen, and Robert~A Santoro.
\newblock Kinetic theory of low-frequency alfv{\'e}n modes in tokamaks.
\newblock {\em Plasma physics and controlled fusion}, 38(11):2011, 1996.

\bibitem{sugama2006collisionless}
H~Sugama and T-H Watanabe.
\newblock Collisionless damping of geodesic acoustic modes.
\newblock {\em Journal of plasma physics}, 72(6):825--828, 2006.

\bibitem{qiu2008collisionless}
Zhiyong Qiu, Liu Chen, and Fulvio Zonca.
\newblock Collisionless damping of short wavelength geodesic acoustic modes.
\newblock {\em Plasma Physics and Controlled Fusion}, 51(1):012001, 2008.

\bibitem{qiu2010nonlocal}
Zhiyong Qiu, Fulvio Zonca, and Liu Chen.
\newblock Nonlocal theory of energetic-particle-induced geodesic acoustic mode.
\newblock {\em Plasma Physics and Controlled Fusion}, 52(9):095003, 2010.

\bibitem{fu2008energetic}
G~Yu Fu.
\newblock Energetic-particle-induced geodesic acoustic mode.
\newblock {\em Physical review letters}, 101(18):185002, 2008.

\bibitem{chavdarovski2021linear}
I~Chavdarovski, M~Schneller, and A~Biancalani.
\newblock Linear dispersion relation of geodesic acoustic modes driven by
  trapped and circulating energetic particles.
\newblock {\em Journal of Plasma Physics}, 87(4), 2021.

\bibitem{novikau2019implementation}
Ivan Novikau, Alessandro Biancalani, Alberto Bottino, Alessandro Di~Siena,
  Ph~Lauber, Emanuele Poli, Emmanuel Lanti, Laurent Villard, No{\'e} Ohana, and
  Sergio Briguglio.
\newblock Implementation of energy transfer technique in {ORB5} to study
  collisionless wave-particle interactions in phase-space.
\newblock {\em Computer Physics Communications}, page 107032, 2019.

\bibitem{nazikian2008intense}
R~Nazikian, GY~Fu, ME~Austin, HL~Berk, RV~Budny, NN~Gorelenkov, WW~Heidbrink,
  CT~Holcomb, GJ~Kramer, GR~McKee, et~al.
\newblock Intense geodesic acousticlike modes driven by suprathermal ions in a
  tokamak plasma.
\newblock {\em Physical review letters}, 101(18):185001, 2008.

\bibitem{zhiyong2011kinetic}
Qiu Zhiyong, Fulvio Zonca, and Chen Liu.
\newblock Kinetic theories of geodesic acoustic modes: Radial structure, linear
  excitation by energetic particles and nonlinear saturation.
\newblock {\em Plasma Science and Technology}, 13(3):257, 2011.

\bibitem{biancalani2017saturation}
A~Biancalani, I~Chavdarovski, Z~Qiu, A~Bottino, D~Del~Sarto, A~Ghizzo,
  {\"O}zg{\"u}r G{\"u}rcan, Pierre Morel, and I~Novikau.
\newblock Saturation of energetic-particle-driven geodesic acoustic modes due
  to wave--particle nonlinearity.
\newblock {\em Journal of Plasma Physics}, 83(6), 2017.

\bibitem{chavdarovski2009effects}
Ilija Chavdarovski and Fulvio Zonca.
\newblock Effects of trapped particle dynamics on the structures of a
  low-frequency shear alfv{\'e}n continuous spectrum.
\newblock {\em Plasma Physics and Controlled Fusion}, 51(11):115001, 2009.

\bibitem{chavdarovski2014analytic}
Ilija Chavdarovski and Fulvio Zonca.
\newblock Analytic studies of dispersive properties of shear alfv{\'e}n and
  acoustic wave spectra in tokamaks.
\newblock {\em Physics of Plasmas}, 21(5):052506, 2014.

\bibitem{cao2015fast}
Jintao Cao, Zhiyong Qiu, and Fulvio Zonca.
\newblock Fast excitation of geodesic acoustic mode by energetic particle
  beams.
\newblock {\em Physics of Plasmas}, 22(12):124505, 2015.

\bibitem{ido2015identification}
T~Ido, M~Osakabe, A~Shimizu, T~Watari, M~Nishiura, K~Toi, K~Ogawa, K~Itoh,
  I~Yamada, R~Yasuhara, et~al.
\newblock Identification of the energetic-particle driven gam in the lhd.
\newblock {\em Nuclear Fusion}, 55(8):083024, 2015.

\bibitem{doi:10.1063/1.5142802}
I.~Novikau, A.~Biancalani, A.~Bottino, Ph. Lauber, E.~Poli, P.~Manz, G.~D.
  Conway, A.~Di~Siena, N.~Ohana, E.~Lanti, and L.~Villard.
\newblock Nonlinear dynamics of energetic-particle driven geodesic acoustic
  modes in asdex upgrade.
\newblock {\em Physics of Plasmas}, 27(4):042512, 2020.

\bibitem{zarzoso2014analytic}
D~Zarzoso, A~Biancalani, A~Bottino, Ph~Lauber, E~Poli, J-B Girardo, X~Garbet,
  and RJ~Dumont.
\newblock Analytic dispersion relation of energetic particle driven geodesic
  acoustic modes and simulations with nemorb.
\newblock {\em Nuclear Fusion}, 54(10):103006, 2014.

\bibitem{girardo2014relation}
Jean-Baptiste Girardo, David Zarzoso, R{\'e}mi Dumont, Xavier Garbet, Yanick
  Sarazin, and Sergei Sharapov.
\newblock Relation between energetic and standard geodesic acoustic modes.
\newblock {\em Physics of Plasmas}, 21(9):092507, 2014.

\bibitem{zarzoso2012fully}
D~Zarzoso, X~Garbet, Y~Sarazin, R~Dumont, and V~Grandgirard.
\newblock Fully kinetic description of the linear excitation and nonlinear
  saturation of fast-ion-driven geodesic acoustic mode instability.
\newblock {\em Physics of Plasmas}, 19(2):022102, 2012.

\bibitem{di2018effect}
A~Di~Siena, A~Biancalani, T~G{\"o}rler, H~Doerk, I~Novikau, P~Lauber,
  A~Bottino, E~Poli, et~al.
\newblock Effect of elongation on energetic particle-induced geodesic acoustic
  mode.
\newblock {\em Nuclear Fusion}, 58(10):106014, 2018.

\bibitem{biancalani2017cross}
A~Biancalani, A~Bottino, C~Ehrlacher, Virginie Grandgirard, G~Merlo, I~Novikau,
  Z~Qiu, E~Sonnendr{\"u}cker, X~Garbet, T~G{\"o}rler, et~al.
\newblock Cross-code gyrokinetic verification and benchmark on the linear
  collisionless dynamics of the geodesic acoustic mode.
\newblock {\em Physics of Plasmas}, 24(6):062512, 2017.

\bibitem{gao2010plasma}
Zhe Gao.
\newblock Plasma shaping effects on the geodesic acoustic mode in the large
  orbit drift width limit.
\newblock {\em Physics of Plasmas}, 17(9):092503, 2010.

\bibitem{betti1992stability}
Ricardo Betti and Jeffrey~P Freidberg.
\newblock Stability of alfv{\'e}n gap modes in burning plasmas.
\newblock {\em Physics of Fluids B: Plasma Physics}, 4(6):1465--1474, 1992.

\bibitem{todo2019introduction}
Y~Todo.
\newblock Introduction to the interaction between energetic particles and
  alfven eigenmodes in toroidal plasmas.
\newblock {\em Reviews of Modern Plasma Physics}, 3(1):1--33, 2019.

\bibitem{biancalani2018nonlinear}
A~Biancalani, N~Carlevaro, A~Bottino, G~Montani, and Z~Qiu.
\newblock Nonlinear velocity redistribution caused by energetic-particle-driven
  geodesic acoustic modes, mapped with the beam-plasma system.
\newblock {\em Journal of Plasma Physics}, 84(6), 2018.

\bibitem{lanti2020orb5}
Emmanuel Lanti, No{\'e} Ohana, Natalia Tronko, Thomas Hayward-Schneider,
  Alberto Bottino, BF~McMillan, Alexey Mishchenko, Aaron Scheinberg, Alessandro
  Biancalani, P~Angelino, et~al.
\newblock Orb5: a global electromagnetic gyrokinetic code using the pic
  approach in toroidal geometry.
\newblock {\em Computer Physics Communications}, 251:107072, 2020.

\bibitem{Weiland_2018}
M.~Weiland, R.~Bilato, R.~Dux, B.~Geiger, A.~Lebschy, F.~Felici, R.~Fischer,
  D.~Rittich, M.~van Zeeland, and and.
\newblock {RABBIT}: Real-time simulation of the {NBI} fast-ion distribution.
\newblock {\em Nuclear Fusion}, 58(8):082032, jul 2018.

\bibitem{vannini2020gyrokinetic}
F~Vannini, A~Biancalani, A~Bottino, T~Hayward-Schneider, Ph~Lauber,
  A~Mishchenko, I~Novikau, E~Poli, and ASDEX~Upgrade Team.
\newblock Gyrokinetic investigation of the damping channels of alfv{\'e}n modes
  in asdex upgrade.
\newblock {\em Physics of Plasmas}, 27(4):042501, 2020.

\bibitem{lauber2014off}
Ph~Lauber.
\newblock Off-axis nbi heated discharges at asdex upgrade: Egams, rsaes, tae
  bursts.
\newblock In {\em 13th Energetic Particle Physics TG Meeting}, 2014.

\bibitem{tronko2016second}
Natalia Tronko, Alberto Bottino, and Eric Sonnendr{\"u}cker.
\newblock Second order gyrokinetic theory for particle-in-cell codes.
\newblock {\em Physics of Plasmas}, 23(8):082505, 2016.

\bibitem{tronko2017hierarchy}
Natalia Tronko, Alberto Bottino, Cristel Chandre, and Eric Sonnendruecker.
\newblock Hierarchy of second order gyrokinetic hamiltonian models for
  particle-in-cell codes.
\newblock {\em Plasma Physics and Controlled Fusion}, 59(6):064008, 2017.

\bibitem{vannini2021studies}
A.~Bottino T. Hayward-Schneider Ph. Laube A. Mishchenko E. Poli B. Rettino G.
  Vlad X.~Wang F.~Vannini, A.~Biancalani and the ASDEX Upgrade~team.
\newblock Studies of the alfvén mode activity in asdex upgrade using the
  isotropic slowing-down distribution function implemented in orb5.
\newblock {\em to be submitted}, 2021.

\bibitem{Laubweb}
Ph. Lauber.
\newblock \url{https://pwl.home.ipp.mpg.de/NLED\_AUG//data.htmll}.

\bibitem{BottinoJPP2015}
A.~Bottino and E.~Sonnendruecker.
\newblock Monte carlo particle-in-cell methods for the simulation of the
  vlasovâ€“maxwell gyrokinetic equations.
\newblock {\em Journal of Plasma Physics}, 81(5):435810501, 2015.

\bibitem{chen1987waves}
Liu Chen.
\newblock {\em Waves and instabilities in plasmas}, volume~12.
\newblock World scientific, 1987.

\bibitem{zonca2008radial}
Fulvio Zonca and Liu Chen.
\newblock Radial structures and nonlinear excitation of geodesic acoustic
  modes.
\newblock {\em EPL (Europhysics Letters)}, 83(3):35001, 2008.

\bibitem{qiu2016effects}
Zhiyong Qiu, Liu Chen, and Fulvio Zonca.
\newblock Effects of energetic particles on zonal flow generation by toroidal
  alfv{\'e}n eigenmode.
\newblock {\em Physics of Plasmas}, 23(9):090702, 2016.

\bibitem{vannini2021gyrokinetic}
F~Vannini, A~Biancalani, A~Bottino, T~Hayward-Schneider, Ph~Lauber,
  A~Mishchenko, E~Poli, G~Vlad, and ASDEX~Upgrade Team.
\newblock Gyrokinetic investigation of the nonlinear interaction of alfv{\'e}n
  instabilities and energetic particle-driven geodesic acoustic modes.
\newblock {\em Physics of Plasmas}, 28(7):072504, 2021.

\end{thebibliography}
\end{document}